\documentclass[fleqn,usenatbib]{mnras}

\usepackage{newtxtext,newtxmath}

\usepackage[T1]{fontenc}

\DeclareRobustCommand{\VAN}[3]{#2}
\let\VANthebibliography\thebibliography
\def\thebibliography{\DeclareRobustCommand{\VAN}[3]{##3}\VANthebibliography}

\usepackage{graphicx}
\usepackage{amsmath}	
\usepackage{placeins}

\newcommand{\msun}{{\rm M}_{\odot}}

\newcommand{\rsun}{{\rm R}_{\odot}}

\title[Effect of mixing on accretor seismology]{The effect of near-core mixing on rejuvenation and the asteroseismic properties of massive accretors}

\author[Henneco \& Bowman]{
Jan Henneco,$^{1}$\thanks{E-mail: jan.henneco@newcastle.ac.uk}
Dominic M. Bowman$^{1,2}$
\\
$^{1}$School of Mathematics, Statistics and Physics, Newcastle University, Newcastle upon Tyne, NE1 7RU, United Kingdom \\
$^{2}$Institute of Astronomy, KU Leuven, Celestijnenlaan 200D, 3001 Leuven, Belgium
}

\date{Accepted 2026 June 7. Received 2026 May 22; in original form 2026 March 27}

\pubyear{\the\year{}}

\begin{document}
\label{firstpage}
\pagerange{\pageref{firstpage}--\pageref{lastpage}}
\maketitle

\begin{abstract}
The relatively recent revelation of the high occurrence rate of binary interactions, especially in intermediate- and high-mass systems, has prompted multiple investigations into their asteroseismic imprints. The near-core region just outside the convective cores of mass-accreting early-type main-sequence stars in binaries has been theorised to be sensitive to assumptions about mixing (notably semiconvection) and accretion physics. In turn, the predicted asteroseismic properties depend strongly on the physical properties of this near-core region. We explore how robust the previously identified asteroseismic imprints of mass accretion are to changes in semiconvective mixing. Using one-dimensional stellar structure and evolution models, this parameter study shows the dominant effect of convective boundary mixing on rejuvenation and the post-accretion asteroseismic properties. The recovered seismic imprint, largely robust to variations in semiconvective mixing efficiency, changes drastically when convective boundary mixing is not included in the models. We find that the post-accretion thermal relaxation is key in determining the final near-core structure and the asteroseismic imprint of accretion. We reaffirm the potential of Fourier transforms of period spacing patterns to quantify the effects of different near-core mixing and accretion-rate assumptions on asteroseismic signals. Overall, this work highlights the sensitivity of the asteroseismic imprint of accretion not only on stellar structure and evolution modelling assumptions, but also on the accretion physics. The logical next step is to arrive at a more general picture of the asteroseismic imprint of mass transfer by exploring its properties in a multi-dimensional parameter study including single- and binary-star assumptions.   
\end{abstract}


\section{Introduction}

The revolution of asteroseismology has been driven by space-based photometric time series data from missions such as Microvariability and Oscillations of STars \citep[MOST;][]{Walker2003}, Convection, Rotation \& planetary Transits \citep[CoRoT;][]{Auvergne2009}, BRIght Target Explorer \citep[BRITE;][]{WeissWW2014}, \textit{Kepler} \citep{Koch2010}, and the Transiting Exoplanet Survey Satellite \citep[TESS;][]{Ricker2015}. These missions have allowed us to characterise the interiors of stars across a wide variety of evolutionary stages and masses (see \citealt{ChaplinMiglio2013}, \citealt{Hekker2017}, \citealt{Bowman2020}, \citealt{Aerts2021review}, and \citealt{Kurtz2022} for reviews). In combination with robust theoretical predictions, diagnostic tools, and ground-based photometric and spectroscopic observations, these space-based variability measurements allow us to precisely determine the radii, masses, and ages of stars (see e.g. \citealt{Chaplin2014, Metcalfe2014, Mombarg2019}). 

For early-type main-sequence stars, the physics of rotation and mixing are especially important for determining their ultimate evolutionary fate (see \citealt{Johnston2021}). For such stars, asteroseismology utilises powerful diagnostic tools, known as gravity-mode period spacing patterns, for probing these physical processes \citep{Miglio2008, Bouabid2013, Pedersen2018}. Thus, high-precision constraints on the convective core properties, as well as interior rotation and mixing profiles are becoming increasingly available \citep{Briquet2007,Briquet2011,Moravveji2015, Moravveji_etal2016, VanReeth2015b, VanReeth2015a, VanReeth2016,VanReeth2018, Li2019, Li2020, Mombarg2019, Wu2019, Wu2020astero, Michielsen2021, Bowman2021, Pedersen2021}. 

However, the majority of early-type stars are born, evolve, and interact with a companion star during their lives \citep{Sana2012, Moe2017a, Offner2023a}. The near-ubiquity of multiplicity (i.e. binaries and higher-order multiples) for massive stars in particular is a double-edged sword when attempting to constrain stellar structure and evolution theory with asteroseismology. In relatively wide and/or pre-mass-transfer binary systems, the application of Kepler's laws of motion to eclipsing binaries yields model-independent masses and radii, which are tight constraints on stellar evolution models when compared to asteroseismic diagnostics alone (see \citealt{SouthworthBowman2025review} for a review). On the other hand, when the components of a binary system interact, their subsequent evolution deviates notably from that of single stars (see \citealt{Marchant2024a} for a review). It is valuable to use asteroseismology to discern whether or not a binary system has undergone an epoch of interaction. It allows us to exclude post-interaction stars when testing single stellar structure and evolution theory, while post-interaction binary products are key to understanding some of the open questions in binary physics, such as mass transfer efficiency. 

Donor stars that (partially or fully) lose their envelope during mass transfer live on as hot subdwarfs \citep{Heber2016, Gotberg2018}, intermediate-mass stripped stars \citep{Podsiadlowski1992a, Gotberg2018, Ludwig2026}, or Wolf-Rayet (WR) stars \citep{Crowther2007}, depending on their stripped-state mass. The mass-gaining accretor stars potentially rejuvenate \citep[e.g.][]{Neo1977,Hellings1983,Braun1995} and spin up, rendering mass transfer as a compelling channel to form Be stars \citep[e.g.][]{Raguzova2005, Liu2006, Peters2008, Wang2017, Wang2021, Milone2018, Hastings2020, Hastings2021, WangChen2020, WangChen2023}, which can have a range of relative rotation rates \citep{Wang2026}. In both cases, their different structures compared to those of single stars translate into different asteroseismic diagnostics. In more extreme cases, mass transfer can result in stellar mergers (see \citealt{Schneider2025review} for a recent review), which pulsate measurably different from genuine single stars and should, therefore, be accounted for when modelling observed stars and populations \citep{Bellinger2024, Henneco2024b, Henneco2025}. As an additional layer of complexity (and opportunity for future exploration), mergers have been shown to be common in triple systems \citep{Kummer2025}, resulting in binary systems with a merger product component.

The renewed interest in the evolutionary properties of accretor stars, see for example, \citet{Renzo2021}, \citet{Renzo2023}, \citet{Lau2024}, \citet{Schurmann2024}, \citet{Zhao2024}, \citet{Bear2025}, \citet{Richards2025}, \citet{Mukhija2026} and \citet{Wang2026}, has sparked recent investigations into their asteroseismic properties. \citet{Miszuda2021} modelled the $\delta$ Scuti accretor star in an intermediate-mass eclipsing binary system and demonstrated the effect of helium enrichment on its pulsations. Similarly, \citet{Miszuda2022} studied another eclipsing binary and showed the effect of ongoing mass transfer on the $\delta$ Scuti accretor star's radial pulsations. In terms of more evolved binary systems containing low-mass stars, \citet{Deheuvels2022} concluded that mass transfer was responsible for the asteroseismic properties of a peculiar sample of red giants. Again for intermediate-mass main-sequence stars, \citet{Wagg2024} studied the impact of accretion on the pulsations of a $3.5\,\msun$ slowly pulsating B-type star. They found that the accretion-induced changes to the near-core region of the accretor leave a distinct and persistent imprint on its asteroseismic characteristics. This study was later repeated for a high-mass $10\,\msun$ accretor by \citet{Miszuda2025betaCep}, who came to similar conclusions. Finally, \citet{Wu2026} recently studied the imprints of accretion on a larger set of low- and intermediate-mass stars, and conclude that the imprints of accretion should be observable in asteroseismic data.

All of these `accretor seismology' studies highlight that the near-core region of accretors are strongly influenced by the increase in mass, which in turn influences the gravity-mode pulsations. This comes as no surprise, since gravity modes are most sensitive to the physical properties of the near-core regions of main-sequence stars (see \citealt{Aerts2010book}). Importantly, however, what these works also have in common is that they use fixed assumptions for the efficiencies of the key mixing processes in the near-core region; convective boundary mixing (CBM) and semiconvection. We note that \citet{Wagg2024} tested different (minimum) amounts of envelope mixing efficiencies, and found that, although their results are robust against a large range of efficiencies, the asteroseismic imprint of the accretion decreases with higher mixing efficiencies during the post-mass-transfer evolution. Semiconvection can safely be ignored in main-sequence single stars with birth masses roughly above $1.7\,\msun$ \citep{Moore2016} because such stars have receding convective cores (see, e.g. \citealt{Kaiser2020} and \citealt{Sibony2023}). However, semiconvection is critical in setting the amount of convective core growth, and hence rejuvenation, of an accreting main-sequence star \citep{Braun1995}. Given its large uncertainties for massive stars \citep[e.g.][]{Kaiser2020,Sibony2023}, and its predicted strong influence of the evolution of the near-core region of accretors, it is a logical next step to assess how the efficiency of semiconvection impacts the predicted asteroseismic imprints of accretion in main-sequence stars.

In this work, we examine if and how the asteroseismic predictions for massive accretors change with different semiconvective efficiencies. Specifically, we use the intermediate-mass binary system from \citet{Wagg2024} as a starting point. In our parameter study, we aim to show how a variety of accretion induced stellar structures and asteroseismic features can be reached for a reasonable range of semiconvective efficiencies. We begin by detailing the computational setup for our binary evolution and stellar pulsation computations in Sect.~\ref{sec:methods}. In Sect.~\ref{sec:results_standard}, we describe and discuss the effect of varying semiconvection efficiencies on the interior structure of the accretor and the corresponding asteroseismic predictions. To decouple the effects of semiconvection and CBM on our results in Sect.~\ref{sec:results_standard}, we repeat the analysis for models without CBM in Sect.~\ref{sec:results_noos}. Finally, we conclude in Sect.~\ref{sec:conclusions}.

\section{Methodology}\label{sec:methods}
In this work, we use almost the same computational setup and physical assumptions for our stellar structure and evolution models as \citet{Wagg2024}, who studied the asteroseismic imprints of accretion for a representative slowly-pulsating B star. However, the key difference is that we recomputed the model from \citet{Wagg2024} with varying efficiencies for semiconvection. We computed two sets of these models; one set with the same assumptions for CBM as \citet{Wagg2024} and one set without CBM. The reason for this is because CBM is another important mixing process that affects rejuvenation and the near-core structure in accretors (see Sect.~\ref{sec:results_standard}).

\subsection{Stellar structure and evolution models}\label{sect:method:sse}

We used the binary module of the stellar structure and evolution code \texttt{MESA} \citep[\texttt{r23.05.1};][]{Paxton2011,Paxton2013,Paxton2015,Paxton2018,Paxton2019,Jermyn2023}. We assumed no wind mass loss since it is expected to be negligible for such stars, and our models do not include rotation. 

Inspired by \citet{Wagg2024}, we initiate a binary system with a $4\,\msun$ primary star, a $3\,\msun$ secondary star, and an orbital period of 5~d and solar metallicity ($Z=0.02$, $Y=0.28$). The chosen binary configuration leads to a mass transfer phase after the primary star reaches the terminal-age main sequence (TAMS) and expands on the Hertzsprung gap. The ensuing mass transfer phase is one that occurs on the thermal timescale of the donor star, since the latter is out of thermal equilibrium on the Hertzsprung gap. Mass transfer was terminated after the accretor accreted $0.5\,\msun$. The mass transfer efficiency was set to 0.5, which means that the accretor only accretes half of the transferred mass. The other half is assumed to be instantly lost from the system. We used the \texttt{Kolb} mass-transfer scheme \citep{Kolb1990}. Following \citet{Wagg2024}, we did not further evolve the donor (i.e. initial primary) star and treated it as a point mass after mass transfer ceased. We followed the post-accretion evolution of the accretor until the TAMS, which we define as when the central hydrogen mass fraction, $X_{\mathrm{c}}$, reached 0.001. 

We used the Ledoux criterion \citep{Ledoux1947} to determine which regions are unstable to convection, which states that regions are unstable if
\begin{align}
    \label{eq:Ledoux}&\nabla_{\mathrm{rad}} > \nabla_{\mathrm{ad}} + \frac{\varphi}{\delta}\nabla_{\mu}\,,\\
    &\text{with}\quad \nabla_{\mathrm{rad}}=\frac{\mathrm{d}\ln T}{\mathrm{d}\ln P}\,,\quad \nabla_{\mathrm{ad}} = \left(\frac{\mathrm{d}\ln T}{\mathrm{d}\ln P}\right)_{\mathrm{ad}}\,,\quad \nabla_{\mu} = \frac{\mathrm{d}\ln \mu}{\mathrm{d}\ln P}\,, \\
    &\varphi= \frac{\partial\ln \rho}{\partial \ln \mu}\,,\quad\text{and}\quad \delta = -\frac{\partial\ln \rho}{\partial \ln T}\,.
\end{align}
In the expressions above, $T$, $P$, $\mu$, and $\rho$ are the local temperature, pressure, mean molecular weight, and density, respectively, and `ad' stands for `adiabatic'. The Schwarzschild criterion for convection, which does not account for the stabilising effect of the mean molecular weight gradient $\nabla_{\mu}$, reads
\begin{equation}\label{eq:Schw}
    \nabla_{\mathrm{rad}} > \nabla_{\mathrm{ad}} ~ .
\end{equation}
Convection was modelled using the mixing length theory \citep[MLT;][]{Bohm-vitense1958,Cox1968} with an efficiency of $\alpha_{\mathrm{MLT}}=2.0$. We used the exponential overshooting scheme from \citet{Herwig2000} to model CBM above the convective hydrogen-burning core, with efficiency parameters $f_{\mathrm{ov}}=0.010$ and $f_{0}=0.005$, which result in an $f_{\mathrm{CBM}} = 0.005$\footnote{To determine the diffusion coefficient of the overshooting region, the overshooting scheme takes the value of the convective diffusion coefficient at a distance of $f_{0}H_{P}$, with $H_{P} = -P(\mathrm{d}r/\mathrm{d}P)$ the pressure scale height, into the convective region. Starting from this point, the scheme then effectively extends the convective region by $f_{\mathrm{ov}}H_{P}$ beyond this point, either with an exponentially-declining or step-like profile in the diffusion coefficient. The effective extent of the overshooting region is therefore dictated by $f_{\mathrm{CBM}} = f_{\mathrm{ov}} - f_{0}$.}. 

Regions which are stable to convection following the Ledoux criterion and unstable according the the Schwarzschild criterion experience semiconvective mixing. We treated semiconvection through the default \citet{Langer1985} scheme in \texttt{MESA r23.05.1}. The base model from \citet{Wagg2024} assumed an efficiency of $\alpha_{\mathrm{sc}}=10^{-1}$ for semiconvection. For our parameter study, we adopted values of $\alpha_{\mathrm{sc}}= 0.0, 10^{-3}, 10^{-2}, 10^{-1}, 10^{0}, 10^{1}, 10^{2}, 10^{3}$, as well as a model in which the Schwarzschild condition is used instead of the Ledoux condition for convection. In light of this parameter study, one can think of models using the Schwarzschild criterion as having infinite semiconvective efficiency \citep{Braun1995}. On the contrary, for $\alpha_{\mathrm{sc}} = 0$, there is no semiconvection. In regions with no convective, semiconvective or convective boundary (i.e. overshooting) mixing, we adopted a constant diffusive mixing coefficient of $D_{\mathrm{min}} = 20\,\mathrm{cm}^{2}\mathrm{s}^{-1}$. 

For comparison purposes, we also computed a suite of single star models to ascertain differences to the accretor models. The masses of these single-star models were chosen in a mass range of $3.45$ to $3.55\,\msun$ around the final accretor mass of $3.50\,\msun$. This was done to cover the Hertzsprung--Russell (HR) diagram evolution of the different accretor models, which do not always coincide with that of a $3.50\,\msun$ single-star model. All single-star models were computed with a semiconvective efficiency of $\alpha_{\mathrm{sc}}=0.1$, though this efficiency does not affect the main-sequence evolution of stars with receding convective cores of that mass in any way.

Lastly, we emphasise that we completed convergence tests based on the effective temperature $T_{\mathrm{eff}}$ and luminosity $L$ to ensure sufficiently high enough spatial and temporal resolution was used in our \texttt{MESA} models. More precisely, we use 10 times more temporal resolution, and a \texttt{mesh\_delta\_coeff} of 0.075 for the spatial resolution instead of 0.4 (for which a lower value means higher resolution) used in \citet{Wagg2024}. In practice, this means that our models are not one-to-one identical to those in \citet{Wagg2024} because ours have much higher resolution, yet they are similar enough for all intents and purposes. Our complete \texttt{MESA} setup is available online.\footnote{\url{https://doi.org/10.5281/zenodo.20623400}}
 
\subsection{Stellar pulsation calculations}\label{sect:method:pulsations}
We used stellar structure profiles of the accretor and single-star \texttt{MESA} models at specific evolutionary ages, proxied by $X_{\mathrm{c}}$, as input for calculations with the stellar pulsation code \texttt{GYRE} \citep[\texttt{v7.1;}][]{Townsend2013,Townsend2018,Goldstein2020,Sun2023} to predict the pulsation frequencies. We computed the non-rotating pulsation frequencies for zonal dipole gravity modes (i.e. $(\ell,\,m) = (1,\,0)$, with $\ell$ and $m$ being the spherical degree and azimuthal order, respectively). We solved the oscillation equations in the adiabatic regime using the \texttt{MAGNUS\_GL6} solver, with vacuum outer boundary conditions. Our frequency scan range and sampling points were slightly different from those of \citet{Wagg2024}: we used 4000 instead of 2000 points and a variable frequency $f$ range that captures modes with radial orders $n$ between 1 and 200 instead of $f \in [0.25;\,10]\,\mathrm{d}^{-1}$. We found that this difference in sampling does not change our results. A direct comparison between the pulsation frequencies recovered for the same stellar structure profile but for these two different frequency ranges and number of sampling points shows that they are virtually identical, with the differences in mode periods being smaller than a few seconds in the worst cases.

\section{Effect of semiconvection on rejuvenation and seismic predictions}\label{sec:results_standard}

\begin{figure*}
\begin{centering}
\includegraphics[width=0.95\textwidth]{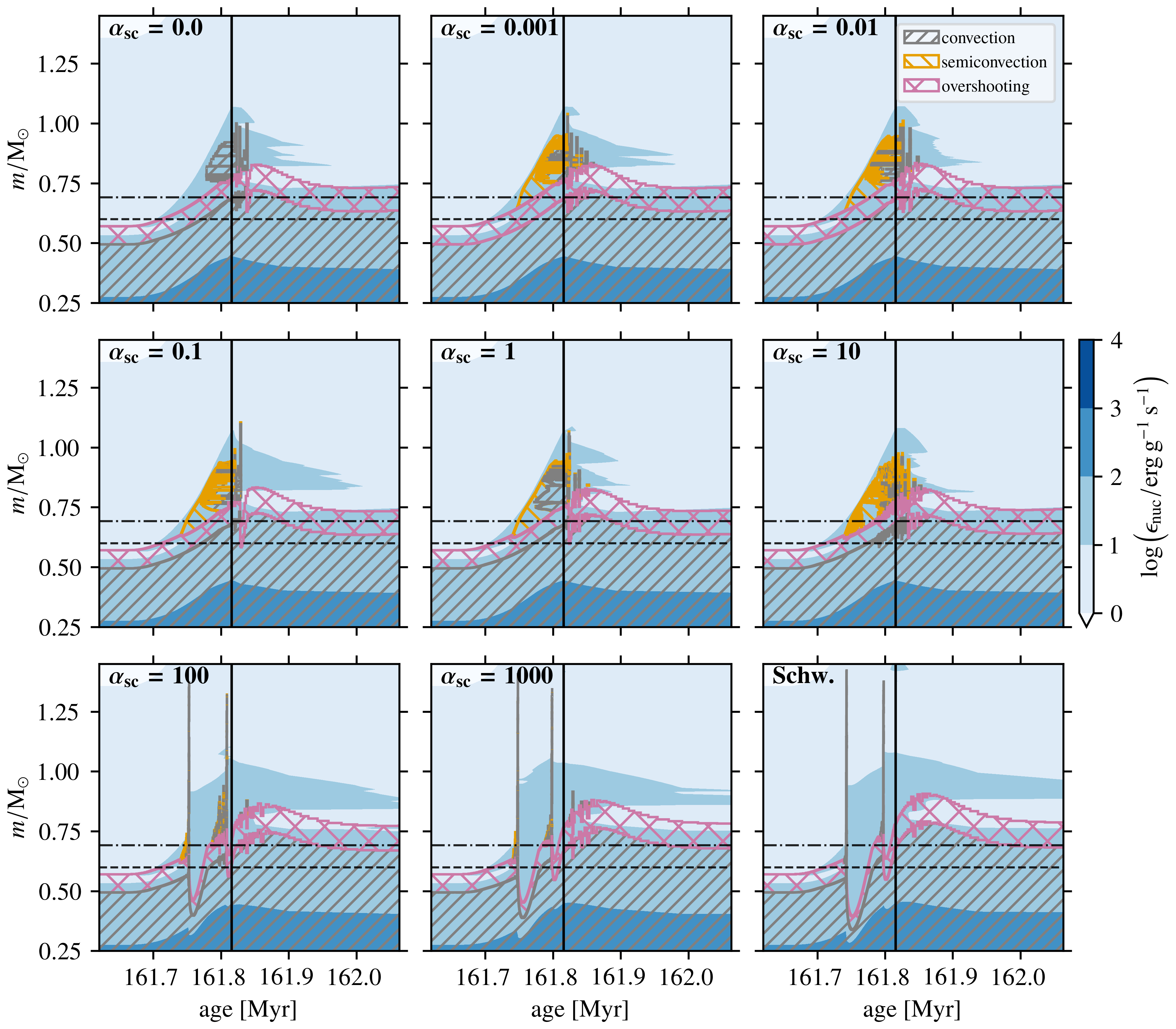}
\par\end{centering}
    \caption{Kippenhahn diagrams of the near-core region for a $3.0+0.5\,\msun$ accretor star with different values of semiconvective mixing efficiency, $\alpha_{\mathrm{sc}}$. `Schw.' stands for `Schwarzschild' and indicates the only model that uses the Schwarzschild criterion for convection instead of the Ledoux criterion. Each Kippenhahn diagram is shown from the onset of mass transfer at $t = 161.62\,\mathrm{Myr}$ until roughly one global thermal timescale $\tau_{\mathrm{KH}} \approx 0.25\,\mathrm{Myr}$ (for the accretor star) after mass transfer ends. The end of mass transfer is indicated by the solid black vertical line. The dashed horizontal black lines indicate the largest pre-mass-transfer extent of the convective core, whereas the dash-dot lines indicate the same but for the combined convective and overshooting region. Part of the stellar core and envelope are left out for clarity.}
    \label{fig:all_kipps_asc}
\end{figure*}

\begin{figure}
 \includegraphics[width=\columnwidth]{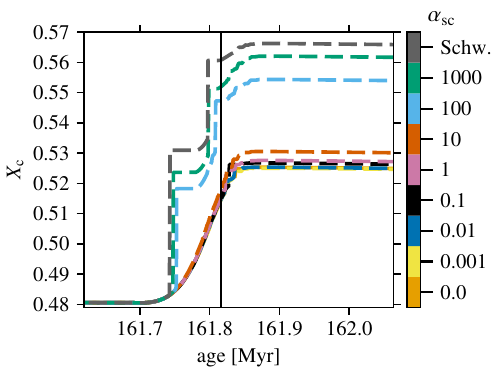}
 \caption{Central hydrogen mass fraction, $X_{\mathrm{c}}$, evolution in time, shown between the onset of mass transfer until one global thermal timescale of the accretors with overshooting after the end of mass transfer. The model from \citet{Wagg2024} with $\alpha_{\mathrm{sc}} = 0.1$ is shown in black. The solid black vertical line indicates the end of mass transfer.}
 \label{fig:center_h1_asc}
\end{figure}

In Fig.~\ref{fig:all_kipps_asc}, we show Kippenhahn diagrams for all accretor models for different semiconvective efficiencies specified in Sect.~\ref{sect:method:sse} and CBM. The Kippenhahn diagrams show the interior mixing and nuclear burning evolution from the moment mass transfer starts until roughly one global thermal timescale, which is defined as $\tau_{\mathrm{KH}} = GM^{2}/(2RL)$, after mass transfer has ceased. In this expression for $\tau_{\mathrm{KH}}$, $M$, $R$, and $L$ are the stellar mass, radius, and luminosity, respectively, and $G$ is the gravitational constant. The duration of the mass-transfer phase is defined by the mass-transfer rate $\log(\dot{M}_{\mathrm{trans}}/\msun\mathrm{yr}^{-1}) \geq -16$. To keep the time-axes identical for all models, we use $\tau_{\mathrm{KH}}$ of the $\alpha_{\mathrm{sc}}=0.1$ model for all Kippenhahn diagrams in Fig.~\ref{fig:all_kipps_asc}. For reference, this thermal timescale $\tau_{\mathrm{KH}} \simeq 1\,\mathrm{Myr}$ for the $3.5\,\msun$ accretor.

Because of the increase in total mass, from $3.0\,\msun$ to $3.5\,\msun$, the accretor's central temperature increases, leading to an increase in the nuclear burning rate and an increase in core luminosity. This causes the accretor's temperature structure to change. As a result, the near-core region previously stable to convection now become Schwarzschild unstable (Eq.~\ref{eq:Schw}); the convective core attempts to grow to a size appropriate for the star's new mass. For models using the Ledoux criterion for convective stability (Eq.~\ref{eq:Ledoux}), which are all models except the one labelled `Schw.' for `Schwarzschild' in Fig.~\ref{fig:all_kipps_asc}, the mean molecular weight gradient built up around the receding convective core prior to mass transfer should prevent the convective core from growing into this region. This region, which is stable to convection according to the Ledoux criterion and unstable according to the Schwarzschild criterion has semiconvective mixing. This can be seen in the Kippenhahn diagrams for all models except the one using the Schwarzschild criterion and the one with $\alpha_{\mathrm{sc}}=0.0$ in Fig.~\ref{fig:all_kipps_asc}. With varying degrees of efficiency, this semiconvective mixing erases the mean molecular weight gradient, allowing the convective core to grow and the accretor to rejuvenate.

The above description is at least the classical picture as stated by \citet{Braun1995}. Upon closer inspection, however, we see that despite the barrier introduced from the chemical gradient in the near-core region, the convective cores of all models grow in mass. This means that another process is enabling the convective core growth, and this process is more efficient in doing so than semiconvection, especially for models with $\alpha_{\mathrm{sc}} \leq 10$. Even before the near-core region becomes unstable to convection according to the Schwarzschild criterion and semiconvection appears, we clearly see in Fig.~\ref{fig:all_kipps_asc} that the convective core considerably grows in mass after the onset of mass transfer. The process responsible for the growth of the convective region that was not included in the original study by \citet{Braun1995} is CBM, which we have implemented as an exponential diffusive profile with $f_{\mathrm{CBM}} = 0.005$ in our models (see Sect.~\ref{sect:method:sse}). The exponential CBM prescription from \citet{Herwig2000} effectively extends the mixing region of the convective core into the chemically stratified near-core region. The extent of the exponential overshooting region depends on the pressure scale height, $H_{P} = -P(\mathrm{d}r/\mathrm{d}P)$, at the convective boundary \citep{Herwig2000}. Thus, it is not directly affected by the mean molecular weight gradient in the near-core region. As a result, overshooting can progressively erase the mean molecular weight gradient in the near-core region, allowing the convective core to grow after mass transfer has begun. In other words, CBM allows accretors to rejuvenate, even without the presence of semiconvection, which is an important difference with the commonly adopted picture from \citet{Braun1995}. This CBM-driven rejuvenation can also be seen in the model from \citet{Miszuda2025betaCep}, who use $\alpha_{\mathrm{sc}}=0.1$ and a larger exponential CBM efficiency of $f_{\mathrm{CBM}} = 0.019$ ($f_{\mathrm{ov}} = 0.020$ and $f_{0} = 0.001$), but did not vary semiconvection. We emphasize that even in the case of infinitely efficient semiconvection, which is the model using the Schwarzschild condition for convection, CBM is an important facilitator of convective core growth.

Thanks to the horizontal lines in Fig.~\ref{fig:all_kipps_asc}, which indicate the largest extent of the pre-mass-transfer convective and overshooting regions, respectively, we see that semiconvection appears well above this region. This is true even before the convective shells appear and influence the chemical composition profile (see Sect.~\ref{sec:convective_shells}). Despite these layers being `untouched' by convection or overshooting, closer inspection shows that they are not chemically homogeneous. Outside the convective core the pp-chain for nuclear hydrogen burning is active. Due to its temperature sensitivity, this leads to a non-negligible mean molecular weight gradient that is enough to stabilise the region against convection according to the Ledoux criterion. The accretion-induced rise in temperature gradient throughout the whole star triggers the appearance of semiconvection in these layers.

\subsection{Accretion-induced convective shells}\label{sec:convective_shells}

The next important feature in the Kippenhahn diagrams in Fig.~\ref{fig:all_kipps_asc} is the development of one or more convective shells, which have been called off-centre convection zones \citep[oCZs;][]{Miszuda2025off}. They have lifetimes of $\sim 10^{5}$ years and occur in all models with $\alpha_{\mathrm{sc}} \leq 10$. These convective shells only appear above the region containing the chemical composition gradient left behind from the receding convective core before mass transfer began. The maximum mass of the pre-mass-transfer convective core is $0.60\,\msun$. Taking into account the contribution of overshooting, which effectively extends the convective core, for the development of the chemical composition gradient, the upper mass coordinate of this region is $0.69\,\msun$. Above this mass coordinate, the convective zones develop quasi-uninhibited by the Ledoux criterion's dependence on the mean molecular weight gradient $\nabla_{\mu}$. Subsequently, they cannot expand to the layers below because of the Ledoux criterion. Despite this, the convective shells can contribute somewhat to rejuvenation, even though they never formally merge with the growing convective core. The convective shell homogenises part of the pp-chain-induced $\mu$ profile. In this way, it deposits hydrogen-rich material in the layers eventually enveloped by the growing convective core (including the CBM region on top). Thus, the convective shell(s) contribute to rejuvenation, albeit marginally compared to the CBM-driven and semiconvection-driven regions.

These findings are mostly in line with what was reported in \citet{Miszuda2025off} and \citet{Miszuda2025betaCep} regarding the appearance and evolution of so-called off-centre convection zones. However, we find that most of the impact of the convective shells on the chemical structure of the near-core region is eventually washed out by the growing convective core and CBM region, which ultimately dictate the chemical structure once the accretor re-commences its main-sequence evolution after mass transfer has ceased. Chemical composition features in layers above the maximum extent of the rejuvenated core are relatively quickly smoothed out by even low to moderate amounts of envelope mixing ($D_{\mathrm{min}}$ in our models).

Next, we see in Fig.~\ref{fig:all_kipps_asc} that the evolution of accretor models with $\alpha_{\mathrm{sc}} \geq 100$ and the one using the Schwarzschild criterion for convection is qualitatively different from those with lower semiconvective efficiencies. Instead of the occurrence of semiconvection on top of the overshooting region and later the development of a convective shell, we find that these models experience multiple episodes with extended, short-lived convective shells accompanied by rapid recession of the convective core and diminishing nuclear burning rates in the core. We also note that because of the efficient chemical homogenisation by semiconvection, these convective shells can connect to the convective core. In between these episodes, the core recovers again and the convective shells disappear. Close inspection shows that each model experiences two large episodes and multiple smaller episodes, which are more noticeable before the second large episode. Similar to the lower semiconvective efficiency models, the accretors have a final convective core growth phase after mass transfer ceases. The occurrence of these episodes can be explained as follows: a large step-like feature appears in the previously smooth mean molecular weight $\mu$ profile because of the efficient semiconvective mixing. Moreover, since the semiconvective region at these efficiencies is fragmented, multiple large discrete steps between the multiple homogeneously mixed shells appear. The same effect is achieved by the fragmented convection that appears after some time in the $\alpha_{\mathrm{sc}}=100$ and $\alpha_{\mathrm{sc}}=1000$ models, and immediately in the Schwarzschild model. These sudden drops in $\mu$ cause sudden jumps in the opacity $\kappa$, which act as radiative flux barriers, since the radiative flux $F_{\mathrm{rad}} \propto \kappa^{-1}$. The build-up of a temperature gradient below these flux barriers eventually gives rise to the extended convective shells, until the process repeats again. This process is reminiscent of the heat-engine $\kappa$-mechanism responsible for exciting stellar oscillations in early-type stars (see e.g. \citealt{Dziembowski1993a, Dziembowski1993b}), though it acts on secular timescales in our case. Recomputing the models with bursts at 4 times higher and lower temporal resolution did not alter the occurrence and appearance of the bursts. This indicates that the origin of these bursts are likely not numerical. We emphasise that the step-like features in the $\mu$ profile are a result of sharp discontinuities in the mixing coefficient $D_{\mathrm{mix}}$ between layers with efficient (i.e. convective, semiconvective, and overshooting) mixing and those with inefficient mixing (i.e. the assumed $D_{\mathrm{min}}$ for the envelope). Therefore, the heat-engine-like behaviour and the associated rejuvenation could very well be consequences of the implementation of mixing in our models. To establish whether or not this mechanism is a consequence of the mixing implementation, one could attempt to replicate it in other stellar evolution codes with different mixing prescriptions. Another way could be to implement alternative mixing prescriptions including the smoothing of diffusion coefficients between different mixing regions in the current models. Both of these avenues are outside the scope of the current work. \\

\subsection{Rejuvenation in numbers}

\begin{table}
\centering
\caption{Helium (He) core masses at TAMS of the accretor and single-star models.}
\label{tab:He_core_masses}
\begin{tabular}{lc@{\hspace{5em}}lc}
\hline\hline
$\alpha_{\mathrm{sc}}$ & $M_{\mathrm{He}}^{\mathrm{TAMS}}$ & $M_{\mathrm{single}}$  & $M_{\mathrm{He}}^{\mathrm{TAMS}}$ \\
                       & $[\mathrm{M}_{\odot}]$            & $[\mathrm{M}_{\odot}]$ & $[\mathrm{M}_{\odot}]$            \\ \hline
0.0                    & 0.393                             & 3.47                   & 0.388                             \\
0.001                  & 0.393                             & 3.48                   & 0.390                             \\
0.01                   & 0.393                             & 3.49                   & 0.391                             \\
0.1                    & 0.393                             & 3.50                   & 0.393                             \\
1                      & 0.394                             & 3.51                   & 0.394                             \\
10                     & 0.395                             & 3.52                   & 0.396                             \\
100                    & 0.400                             & 3.53                   & 0.397                             \\
1000                   & 0.402                             & 3.54                   & 0.398                             \\
Schw.                  & 0.402                             & 3.55                   & 0.400                               \\
\hline\hline
\end{tabular}
\end{table}

In Fig.~\ref{fig:center_h1_asc} we see the evolution of the central hydrogen fraction of all accretor models shown in Fig.~\ref{fig:all_kipps_asc}. It is clear that all accretor models, including the model with $\alpha_{\mathrm{sc}} = 0.0$, rejuvenate to varying degrees. At the onset of mass transfer the accretor has $X_{\mathrm{c}}=0.48$. Models with $\alpha_{\mathrm{sc}} \leq 10$ rejuvenate to $X_{\mathrm{c}} \simeq 0.52\text{--}0.53$. As expected, more efficient semiconvection leads to more rejuvenation, yet the bulk of rejuvenation is driven by CBM. At the highest semiconvective efficiencies and in the model using the Schwarzschild criterion, the amount of rejuvenation is notably higher. In these models, $X_{\mathrm{c}}$ increases to 0.54, 0.55, and 0.56 for $\alpha_{\mathrm{sc}} = 100$, $\alpha_{\mathrm{sc}} = 1000$, and the Schwarzschild criterion, respectively. In addition to initial progressive rejuvenation, they rejuvenate in two large and several smaller bursts. These rejuvenation bursts are linked to the short-lived extended convection zones caused by the heat-engine-like mechanism described in Sect.~\ref{sec:convective_shells}.

It is clear that a non-negligible part of the rejuvenation process, that is, the convective core mass growth and the associated rise in central hydrogen mass fraction, happens after the mass transfer phase. This can be seen in Figs.~\ref{fig:all_kipps_asc} and \ref{fig:center_h1_asc} thanks to the solid vertical lines that indicate the end of mass transfer for each model. In addition to continued core growth, we see the development of extended regions with one or multiple short-lived convection shells after accretion ceases in models with $\alpha_{\mathrm{sc}}\leq 1$. We attribute both phenomena to the thermal relaxation of the accretor, and elaborate on them and their effects on the final post-mass-transfer structure in Sect.~\ref{sec:peak_origins}.

Overall, we find that the effect of semiconvection on rejuvenation is relatively minor compared to the rejuvenation driven by CBM, as can be judged by the relatively small variation in final $X_{\mathrm{c}}$ values for models with $\alpha_{\mathrm{sc}} \leq 10$. Similar to and more effectively than the convective shells triggered by accretion, semiconvection increases rejuvenation by mixing fresh hydrogen into the region that the convective core eventually expands into. Contrary to the convective shells, semiconvection can operate in the region with the strong chemical composition gradient left behind by the receding convective core. Different semiconvection efficiencies impact the amount of mixing of material from the upper hydrogen-rich layers into the lower regions. For the models that rejuvenate in bursts (i.e. those with $\alpha_{\mathrm{sc}}=100$, $\alpha_{\mathrm{sc}}=1000$, and the Schwarzschild criterion), the effect of semiconvection is more indirect. Since it is not straightforward to verify this from our current models, we speculate that more efficient semiconvection (or convection, in the extreme case of the model using the Schwarzschild condition) leads to sharper step-like profiles in the $\mu$ and $\kappa$ profiles, causing more efficient radiative flux blocking and a more efficient secular heat-engine mechanism to operate. Compared to the rejuvenation during these bursts, the direct (semi-)convection-driven rejuvenation just outside the overshooting region is negligible. Further establishing the intricate workings of this secular heat-engine mechanism are left for future work.

In addition to $X_{\mathrm{c}}$, we define another measure of the amount of rejuvenation, which is the helium core mass at the TAMS $M_{\mathrm{He}}^{\mathrm{TAMS}}$ of the accretor. We provide these values for all models shown in Figs.~\ref{fig:all_kipps_asc} and \ref{fig:center_h1_asc}, and in Table~\ref{tab:He_core_masses}. Comparing these to the values for single-star models in the same table, we see that semiconvection has virtually no effect on the final core mass at TAMS for $\alpha_{\mathrm{sc}} \leq 10$. In fact, using the $M_{\mathrm{He}}^{\mathrm{TAMS}}$ value of the $3.5\,\msun$ single star model as a comparison, we conclude that all models rejuvenate, even those with no or inefficient semi-convection. This further illustrates that CBM is far from negligible relative to semiconvection for rejuvenation.

\subsection{Asteroseismic predictions}\label{sec:asteroseismic_predicitions}

\begin{figure}
 \includegraphics[width=0.95\columnwidth]{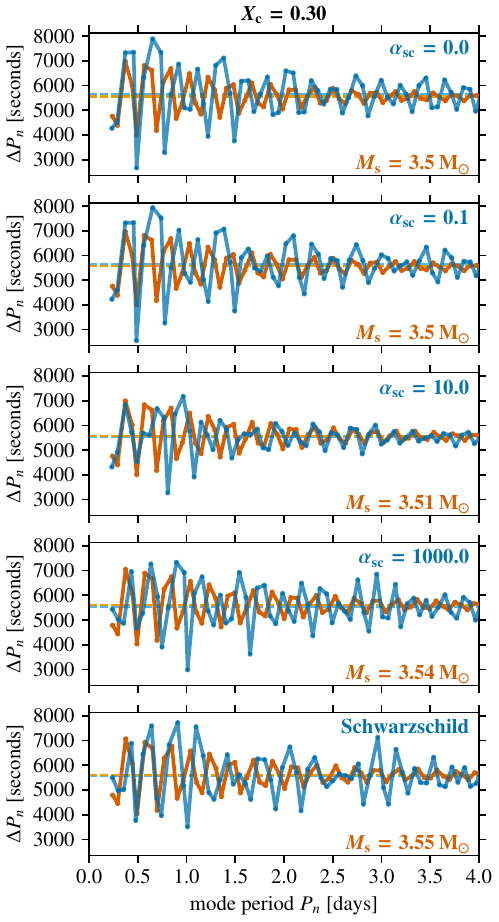}
 \caption{Period spacing patterns (PSPs) for dipole zonal ($\ell = l$, $m=0$) gravity modes without rotation for the $3.5\,\msun$ accretor models (solid blue lines) at $X_{\mathrm{c}} = 0.30$ with different values of $\alpha_{\mathrm{sc}}$ and the model using the Schwarzschild criterion for convection. The solid orange lines represent the PSPs for the equivalent single-star models of mass $M_{\mathrm{s}}$. The dashed blue and orange lines show the asymptotic period spacing $\Pi_{\ell=1}$ for the accretor and single-star models, respectively.}
 \label{fig:PSP_asc_Xc30}
\end{figure}

We calculate the pulsation mode properties for our accretor and equivalent single-star models at different times during the (post-mass-transfer) main-sequence evolution. We do not calculate the pre-mass-transfer pulsations properties of the accretor in this work, since the initial stellar structure and, hence, asteroseismic properties are not affected by varying the semiconvection efficiency. We refer to \citet{Wagg2024} for information on these properties.

In this work, we focus on gravity modes since they are the most common type of pulsations for stars with masses between about 3 and 9~M$_{\odot}$ \citep{Balona2011b, Bowman2019, Burssens2020a}. Moreover, gravity modes are highly sensitive to the physical conditions in the near-core region, thus most affected by varying the semiconvective efficiency. From the pulsation mode periods of the predicted gravity modes, we construct period spacing patterns (PSPs; \citealt{Miglio2008}), which are defined as the differences between the mode periods of modes of consecutive radial order, $n$, and the same spherical degree, $\ell$, and azimuthal order, $m$, as a function of the pulsation mode period, $P_{n}$, such that:
\begin{equation}
    \Delta P_{n} = P_{n} - P_{n-1} ~ .
\end{equation}
PSPs can also be constructed as a function of the radial orders $n$ of the involved pulsation modes, which are used in Sect.~\ref{sec:fourier}.

Figure~\ref{fig:PSP_asc_Xc30} shows the PSPs for the $3.5\,\msun$ accretor model using the Schwarzschild criterion and those with $\alpha_{\mathrm{sc}} = 0.0$, $0.1$, $10$, $1000$ at $X_{\mathrm{c}} = 0.30$. The figure also shows the PSPs of the accretor models' equivalent single-star models. For completeness, the PSPs for models with $X_{\mathrm{c}} = 0.47$, $0.10$, and $0.01$ are shown in Figs.~\ref{fig:PSP_asc_Xc47}--\ref{fig:PSP_asc_Xc1} in Appendix~\ref{app:other_psps}. We specifically show the PSPs for gravity modes of $(\ell,\,m) = (1,\,0)$ without the effects of rotation taken into account.  For each accretor model, the equivalent single-star model is that of a single star with the same $X_{\mathrm{c}}$ and similar effective temperature $T_{\mathrm{eff}}$ and luminosity $L$ as the accretor model. Due to the different amounts of rejuvenation, and hence different convective core masses, not all accretor models have a $3.5\,\msun$ equivalent single-star; more rejuvenated models are closer in $T_{\mathrm{eff}}$ and $L$ to single-star models with higher masses. 

For each PSP shown in Fig.~\ref{fig:PSP_asc_Xc30}, we also show the asymptotic period spacing value, $\Pi_{\ell=1}$, which is the value of the period spacing expected for the high-radial order (i.e asymptotic) regime for a homogeneous, non-rotating, non-magnetic star. The asymptotic period spacing value can be derived directly from a star's equilibrium structure following \citet{Tassoul1980},
\begin{equation}\label{eq:asymptotic_PS}
    \Pi_{\ell} = \frac{\Pi_{0}}{\sqrt{\ell(\ell+1)}}\,,\quad \text{with} \quad \Pi_{0} = \pi\left(\int_{r_{\mathrm{i}}}^{r_{\mathrm{o}}}\frac{\tilde{N}}{r}\mathrm{d}r\right)^{-1} ~ ,
\end{equation}
\noindent where $r$ is the radial coordinate, $r_{\mathrm{i}}$ and $r_{\mathrm{o}}$ the inner and outer radial coordinates of the gravity-mode cavity, and $\tilde{N}$ the linear Brunt-V\"ais\"al\"a (BV) frequency, which is related to its angular version $N$ by $N = 2\pi\tilde{N}$. Since we do not model rotation or magnetic fields on the level of the stellar structure or the pulsation computations, the departures of the PSPs shown in Fig.~\ref{fig:PSP_asc_Xc30} arise solely from the chemical composition gradients in the near-core region of the star.

We emphasise at this point that the PSPs for the $\alpha_{\mathrm{sc}}=0.1$ accretor model and its equivalent single-star model in Fig.~\ref{fig:PSP_asc_Xc30} differ somewhat in terms of the morphology of the quasi-periodic departures from $\Pi_{\ell=1}$ compared to those shown in Fig.~5 of \citet{Wagg2024}. We see this at different ages (Figs.~\ref{fig:PSP_asc_Xc47}--\ref{fig:PSP_asc_Xc1}) as well. This can be explained as follows. First, we use different temporal and spatial resolution in our underlying \texttt{MESA} models. The BV frequency profile in the near-core region, to which gravity modes are extremely sensitive and varies rapidly, is resolved slightly better in our models. Second, we select stellar structure profiles from \texttt{MESA} as input into \texttt{GYRE} at specific $X_{\mathrm{c}}$ values by looking for the profile with a $X_{\mathrm{c}}$ value closest to the desired value of $X_{\mathrm{c}}$. Because we write out profiles every 10 time steps in \texttt{MESA} with 10 times higher temporal resolution than \citet{Wagg2024}, who write out a profile at every time step, there is a slight mismatch in the $X_{\mathrm{c}}$ values of the profiles used by \texttt{GYRE} between our and their work. That said, the goal of the present work is not to compare our \texttt{GYRE} predictions with those of \citet{Wagg2024}, but to explore the effect of near-core mixing and semiconvection, so this small difference is acceptable.

Looking at the asymptotic period spacing values $\Pi_{\ell=1}$ in Fig.~\ref{fig:PSP_asc_Xc30}, we see that the absolute difference in $\Pi_{\ell=1}$ between the accretors and their equivalent single stars is small compared to the overall scatter in $\Delta P_{n}$. In other words, the values of $\Pi_{\ell=1}$ are nearly identical for the accretors and their equivalent single stars. Closer inspection shows that these differences are on the order of several tens of seconds, but never more than 100~seconds. Moreover, we find a relatively small variation in $\Pi_{\ell=1}$ over the range of semiconvective efficiencies shown in Fig.~\ref{fig:PSP_asc_Xc30}, namely $\Pi_{\ell=1} \in [5526;\,5637]\,$s. We find similar ranges at different points along the main sequence, such that $\Pi_{\ell=1} \in [5878;\,6087]\,$s for $X_{\mathrm{c}}=0.47$, $\Pi_{\ell=1} \in [4782;\,4816]\,$s for $X_{\mathrm{c}}=0.10$, and $\Pi_{\ell=1} \in [3728;\,3781]\,$s for $X_{\mathrm{c}}=0.01$. As demonstrated in Fig.~14 of \citet{Henneco2025}, these differences in $\Pi_{\ell=1}$ of $\sim 100\,$s are typically not sufficient to identify stars as outliers in main-sequence populations. 

Next, in line with what was found for the $\alpha_{\mathrm{sc}}=0.1$ model from \citet{Wagg2024}, the higher-mass model from \citet{Miszuda2025betaCep}, and the predictions in \citet{Wu2026}, we observe that for all semiconvective efficiencies, the quasi-periodic variation in the accretors' PSPs has a higher amplitude and is phase-shifted compared to their equivalent single-stars' PSPs. Moreover, whereas the periodicity in the single-stars' PSPs appears somewhat regular, the periodicity of the accretors' PSPs has a second component. We make these differences more quantitative and concrete in Sect.~\ref{sec:fourier} below.

We now zoom out and compare the overall morphology and amplitude of the quasi-periodic PSP variability for the different accretor models. We find that aside from the exact location of specific PSP features, the semiconvective efficiency does not significantly change the overall qualitative behaviour of the PSPs. The model with $\alpha_{\mathrm{sc}}=10$ differs the most from the others in that the amplitude of the PSP variation is similar to those of its equivalent single star. As mentioned in the previous section, the $\alpha_{\mathrm{sc}}=10$ model has the strongest rejuvenation without having the short-lived extended convection zones found for the models with $\alpha_{\mathrm{sc}} \geq 100$ and the model using the Schwarzschild criterion. In other words, from the qualitative assessment of these asteroseismic predictions it appears that the results for the PSPs reported in \citet{Wagg2024} are robust against variations in the semiconvective efficiency. This conclusion can likely be extended to the results of \citet{Miszuda2025betaCep} and \citet{Wu2026}, although a dedicated parameter study at different stellar masses is needed to make this concrete. A large contributor to the lack of strong variations in the PSPs with varying semiconvective efficiencies is the fact that CBM is the main driver for rejuvenation in our models. We explore if and how this picture changes without the effect of overshooting in Sect.~\ref{sec:results_noos}.\\

\begin{figure}
 \includegraphics[width=0.95\columnwidth]{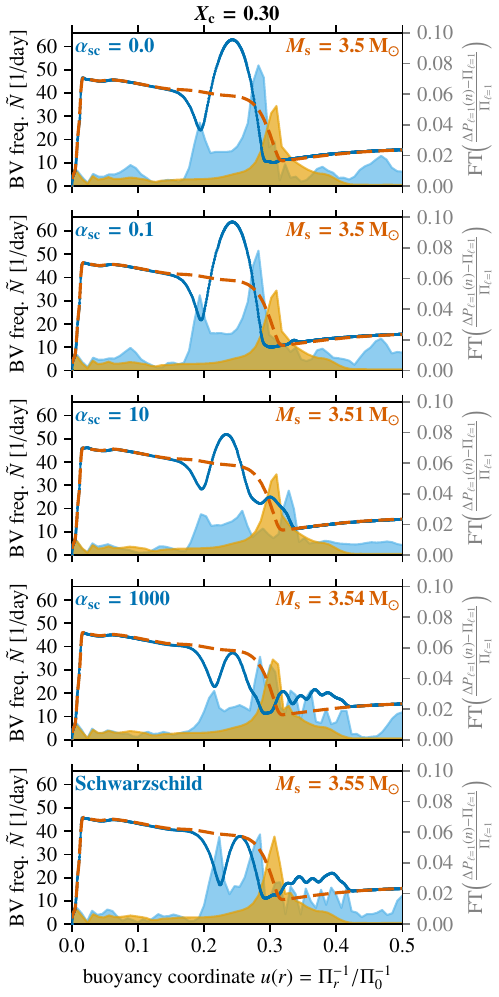}
 \caption{Brunt-V\"ais\"al\"a frequency $\tilde{N}$ profiles (solid blue and dashed orange lines for the accretor and single star, respectively, left y-axis) and Fourier transforms (light blue and gold lines for the accretor and single star, respectively, right y-axis) of the corresponding normalised PSPs for accretors and equivalent single stars as a function of the buoyancy coordinate $u(r)$ for different $\alpha_{\mathrm{sc}}$ at $X_{\mathrm{c}} = 0.30$. The area underneath the Fourier transform is filled for clarity.
 Note that, by definition, the convective core is not shown and the core-envelope boundary is at $u(r) = 0$, and $M_{\mathrm{s}} = M_{\mathrm{single}}$ is the mass of the equivalent single star.}
 \label{fig:BV_and_FT_Xc30}
\end{figure}

Figure~\ref{fig:BV_and_FT_Xc30} shows the BV frequency profiles for the different accretor models and their equivalent single stars shown in Fig.~\ref{fig:PSP_asc_Xc30}. The equivalent figures for $X_{\mathrm{c}} = 0.47$, $0.10$, and $0.01$ can be found in Appendix~\ref{app:other_seismic_figures}. Instead of the radial coordinate $r$ or the mass coordinate $m$, we used the buoyancy coordinate $u(r)$, defined as:
\begin{equation}
    u(r) = \Pi_{r}^{-1}/\Pi_0^{-1} = \dfrac{\int^{r}_{r_{\mathrm{i}}}\left[N(r')/r'\right]\mathrm{d}r'}{\int^{R}_{r_{\mathrm{i}}}\left[N(r')/r'\right]\mathrm{d}r'}\,,
\end{equation}
as the coordinate for the x-axis, with $r_{\mathrm{i}}$ and $r_{\mathrm{o}}$ the radial coordinates of the inner and outer boundaries of the g-mode cavity, respectively \citep{Montgomery2003, Guo2026}. Based on this definition of $u(r)$, it is a natural coordinate to study the BV frequency in the near-core region \citep{Guo2026}.
In the BV frequency profile of the equivalent single stars in Fig.~\ref{fig:BV_and_FT_Xc30} we recognise the typical peak in the near-core region, caused by the chemical composition gradient left behind by the receding convective core during the main sequence. We note that by using $u(r)$ as a coordinate, the convective core is not shown and the core-envelope boundary is at $u(r) = 0$. We find that the additional $\tilde{N}$-peak described in \citet{Wagg2024} and reproduced in our $\alpha_{\mathrm{sc}}=0.1$ model is present in all models, but decreases in magnitude with increasing $\alpha_{\mathrm{sc}}$ (except for the $\alpha_{\mathrm{sc}}=0.0$ model, whose peak is smaller than that of the $\alpha_{\mathrm{sc}}=0.1$ model). To see the reason behind this trend with $\alpha_{\mathrm{sc}}$, we revisit the origin of this additional peak in more detail in Sect.~\ref{sec:effect_on_peak}.

\subsection{The origin of the double-peaked BV frequency profile -- revisited}\label{sec:peak_origins}

\begin{figure*}
\begin{centering}
\includegraphics[width=0.95\textwidth]{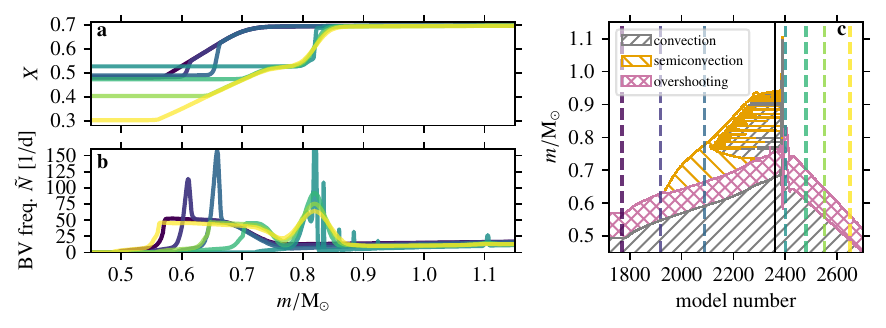}
\par\end{centering}
    \caption{Hydrogen mass fraction $X$ (Panel a), BV frequency $\tilde{N}$ profile (Panel b), and Kippenhahn diagram (Panel c) for the $\alpha_{\mathrm{sc}}=0.1$ accretor model at different times. Each panel only shows the near-core region and contains information from right before the onset of mass transfer until roughly the mid-main sequence of the rejuvenated accretor. The Kippenhahn diagram uses the \texttt{MESA} model number for its x-axis for clarity, since the information it shows spans multiple timescales. The colour of each line in Panels a and b corresponds to the colour of the vertical dashed lines in Panel c. The solid black vertical line indicates the end of mass transfer.}
    \label{fig:double_peak}
\end{figure*}

As seen in the previous section, accretion and the associated rejuvenation enabled by CBM result in the typical double-peaked BV frequency profile feature. Not only is this feature robust against variations in envelope mixing \citep{Wagg2024} and found for a higher-mass accretor \citep{Miszuda2025betaCep}, in this work we demonstrate that it is also robust against variations in $\alpha_{\mathrm{sc}}$. Yet, from the intermediate-mass models of \citet{Wu2026}, it becomes apparent that the exact shape of the double-peaked feature, and therefore also the associated asteroseismic features, can vary. Whilst \citet{Wagg2024} (and by extension our own work) and \citet{Miszuda2025betaCep} recover a $\tilde{N}$-profile with two well-defined separate peaks, the BV frequency profiles in \citet{Wu2026} show a more combined shape, with the accretion-induced peak `attached' to the one caused by the receding convective core. Complementary to the explanations given in the works named above, we investigate the exact nature of these different shapes in this section. In addition to that, this section addresses the fact that the final stages of rejuvenation, which are instrumental in determining the exact location and shape of the accretion-induced peak, occur after the end of mass transfer and so must be driven by another mechanism.

Figure~\ref{fig:double_peak} shows profiles of hydrogen mass fraction, $X$, and BV frequency, $\tilde{N}$, as well as a Kippenhahn diagram spanning right before, during and after mass accretion for the $\alpha_{\mathrm{sc}}=0.1$ model. Right before accretion begins, we see the typical $X$ profile for a mid-main-sequence massive star: a flat $X$ profile for the convective core and the layers mixed by CBM up to $m/\msun = 0.57$, a positive gradient in $X$ in the layers through which the convective core (we treat the convective core and the CBM region on top as the `convective core' in this explanation) receded during the main-sequence evolution at $0.57 < m/\msun  \leq 0.69$, and again a quasi-flat profile for the rest of the envelope (as mentioned in the first part of Sect.~\ref{sec:results_standard}, there is a relatively weak chemical composition gradient due to nuclear burning in the envelope via the pp-chain). During accretion, the convective core becomes larger than the pre-mass-transfer convective core. As a result, the chemical composition gradient between $0.57 < m/\msun \leq 0.69$ eventually becomes fully mixed up to $m/\msun = 0.77$. The connection between the homogeneously mixed convective core and the rest of the stellar envelope causes a relatively sharp chemical composition gradient. The exact shape of this gradient depends on the CBM prescription used (see \citealt{Pedersen2018}). This sharp chemical composition gradient causes the appearance of the outer peak in the BV frequency profile that moves to higher mass coordinates as the core grows, which can be seen in Fig.~\ref{fig:double_peak}b. 

At the location of the solid black line in Fig.~\ref{fig:double_peak}c, mass transfer ceases, yet the convective core and the overshooting region on top grow to a mass coordinate of $m/\msun \simeq 0.83$ for a relatively brief amount of time (see Fig.~\ref{fig:all_kipps_asc} for the same Kippenhahn diagram as a function of time). Simultaneously, a short-lived, extended convective shell appears (closer inspection shows that this is in fact a collection of smaller convective regions not connected to the convective core region). We attribute this to the gravitational energy released by the contraction ($\Delta R \sim 1\,\rsun$) during thermal relaxation after mass transfer ceased \citep{Wang2026}. This contraction occurs on a relatively short timescale, which is a result of the abrupt end of mass transfer. To ensure that this post-mass-transfer core growth is not an artifact of the artificial abrupt end of mass transfer in our models, which is done in both the \citet{Wagg2024} and \citet{Miszuda2025betaCep} models, we recomputed the $\alpha_{\mathrm{sc}}=0.1$ model and decreased the accretion rate progressively between $M_{\mathrm{accretor}} = 3.45\text{--}3.50\,\msun$. From this we found that the convective core reaches the same mass as in the original model during the post-mass-transfer contraction, since the same amount of gravitational energy is released, albeit on a longer timescale. Thus, the contraction phase `pushes' the steep chemical composition gradient responsible for the outer BV frequency peak to a mass coordinate ($m/\msun \simeq 0.83$) beyond where it is right after mass transfer ends ($m/\msun \simeq 0.77$). After the star has fully regained thermal equilibrium, the convective core settles to a mass that is noticeably lower than that of its maximum mass during the contraction phase. Starting from this point, the convective core recedes again, leaving behind a less steep chemical composition gradient, leading to the development of the inner BV frequency peak. The relatively thin plateau in $X$, described in \citet{Wagg2024}, is a direct result of the thermal relaxation of the star, that is the short increase and later decrease of the convective core mass. In other words, the thermal relaxation of the accretor is responsible for the double peak shape of the $\tilde{N}$--profile associated with accretion. This mechanism is similar to the nucleo-thermal relaxation of the core of a main-sequence merger product used to explain the double-peak feature in its BV frequency profile in \citet{Henneco2025}.

\begin{figure*}
\begin{centering}
\includegraphics[width=0.95\textwidth]{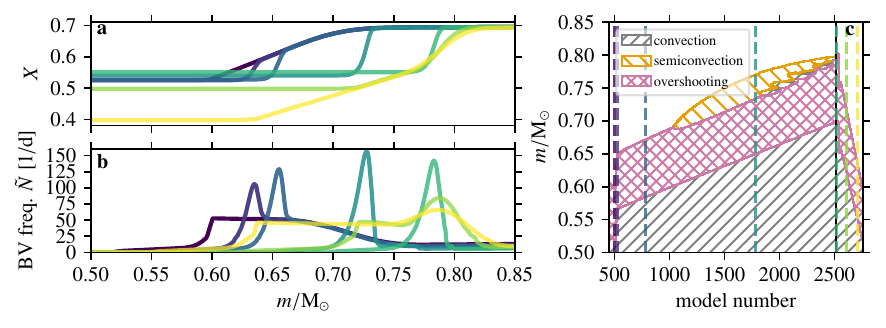}
\par\end{centering}
    \caption{Same as Fig.~\ref{fig:double_peak}, but for an accretor with $\alpha_{\mathrm{sc}}=0.1$ in a binary system with an initial period of 1.5~d instead of 5~d, which results in Case A (i.e. nuclear-timescale) mass transfer.}
    \label{fig:combined_peak}
\end{figure*}

\subsection{Impact of mass accretion rate}

To further substantiate the key role of the thermal relaxation induced contraction on shaping the distinct post-mass-transfer near-core structure of an accretor, we repeated this analysis on an accretor model with (slower) nuclear-timescale mass transfer. We recomputed the base model ($\alpha_{\mathrm{sc}}=0.1$) with an initial orbital period of 1.5~d. This results in nuclear-timescale mass transfer while the donor star is still on the main sequence, whereas the original model experienced thermal-timescale mass transfer. 

Using this lower accretion rate, we show in Fig.~\ref{fig:combined_peak} that the convective core does not increase in mass significantly after the end of mass transfer. This is because the star is in thermal equilibrium during mass transfer and barely contracts when mass transfer ceases ($\Delta R \sim 0.1\,\rsun$). When the core then starts receding during the main-sequence evolution, a new chemical gradient starts building up starting from the bottom of the sharp chemical gradient responsible for the outer BV frequency peak, instead of further down in mass coordinate (as in the original model). In such cases, the resulting chemical composition profile does not have a flat part separating the two gradients, as seen in Fig.~\ref{fig:double_peak}, but instead a single point where the two gradients meet. The resulting BV frequency profile then shows a less clear distinction  of having two BV frequency peaks. 

The shapes of the resulting BV frequency profiles in Fig.~\ref{fig:combined_peak} closely resemble those shown in \citet[][model set A]{Wu2026}, since they model nuclear-timescale mass transfer (with mass transfer rates $\dot{M}_{\mathrm{trans}}\sim 10^{-8}\,\msun\mathrm{yr}^{-1}$). Although \citet{Miszuda2025betaCep} model mass transfer from a MS donor, which generally occurs on a nuclear timescale, they only capture the thermal-timescale mass-transfer phase preceding the nuclear-mass transfer phase. This can be appreciated from the relatively high mass transfer rate ($\dot{M}_{\mathrm{trans}}\sim 10^{-3.5}\,\msun\mathrm{yr}^{-1}$) shown in Fig.~1 from \citet{Miszuda2025betaCep}. As a result, the accretor in their work is also out of thermal equilibrium and contracts with $\Delta R \sim 1\,\rsun$ after the end of mass transfer. This leads to the same transient core growth and shrinkage, and hence the same double-peaked structure in the BV frequency profile, as described in Sect.~\ref{sec:peak_origins}.

We note the following on the accretion-rate argument made above: We consider the fact that the initially $7\,\msun$ accretor from \citet{Miszuda2025betaCep} has a shorter global thermal timescale $\tau_{\mathrm{KH}}$ than the initially $3\,\msun$ accretor from \citet{Wu2026}. This means that the more massive accretor can accommodate larger accretion rates without being out of thermal equilibrium. Therefore, one could argue the comparison between accretion rates is unjustified. Although true in general, we can easily show that our point still holds for the considered masses and accretion rates. If we assume a mass-radius relation $R \propto M^{0.5}$ and a mass-luminosity relation $L \propto M^{3}$ \citep{Kippenhahn2012}, where we ignored the dependence of the mean molecular weight $\mu$ in both relations for the sake of this argument, we find a mass dependence for the thermal timescale of $\tau_{\mathrm{KH}} \propto M^{-3/2}$. With this relation, we find that $\tau_{\mathrm{KH}}^{3\msun} \approx  3.6\times\tau_{\mathrm{KH}}^{7\msun}$. However, despite the fact that the $7\,\msun$ accretor has a roughly 3.6 times shorter thermal timescale than the $3\,\msun$ accretor, its accretion rate is $10^{4.5}$ times larger. In other words, the more massive accretor could in principle handle a mass transfer accretion that is a few times higher than the less massive accretor could, but its accretion rate is orders of magnitude higher.

In this work, we focus on a specific binary system with an initial mass ratio of $q_{\mathrm{i}}=M_{\mathrm{a,i}}/M_{\mathrm{d,i}} = 0.75$, with $M_{\mathrm{d,i}}$ and $M_{\mathrm{a,i}}$ the initial masses of the donor and accretor star, respectively. This experiment should be repeated on a range of different initial binary configurations to explore the different possible BV frequency profile morphologies. Nevertheless, with a quick back-of-the-envelope calculation, we can already estimate which initial mass ratios result in out-of-thermal-equilibrium accretors. For these accretors, we expect to find a similar morphology of the double-peaked BV frequency profile as described in Sect.~\ref{sec:peak_origins} and this section. 

As before, we estimate the global thermal timescale of both components as $\tau_{\mathrm{KH}} = kM^{-3/2}$, with $k$ a constant that depends on $\mu$ that we assume to be the same for both components at the Zero-Age Main Sequence (ZAMS). We assume that the donor star transfers mass on its thermal timescale $\tau_{\mathrm{KH,d}}$. Hence, the mass-transfer rate $\dot{M}_{\mathrm{d}}$ can be expressed as $\dot{M}_{\mathrm{d}}=M_{\mathrm{d}}/\tau_{\mathrm{KH,d}}$. To keep things general, we assume that only a fraction $\beta$ of the transferred mass is accreted, as is the case in our model ($\beta = 0.5$). In other words, the accretion rate $\dot{M}_{\mathrm{a}} = -\beta \dot{M}_{\mathrm{d}}$. Since we are interested in timescales, we only care about the magnitudes of these rates and we can drop the factor $-1$ in this expression. Thus, we can write $\dot{M}_{\mathrm{a}}$ as
\begin{equation}
    \dot{M}_{\mathrm{a}} = \beta \dfrac{M_{\mathrm{d}}}{\tau_{\mathrm{KH,d}}}\quad.
\end{equation}
Using the expression of accretion rate $\dot{M}_{\mathrm{a}}$, the accretion timescale $\tau_{\mathrm{accr}}$ can written as
\begin{equation}
    \tau_{\mathrm{accr}} = \dfrac{M_{\mathrm{a}}}{\dot{M}_{\mathrm{a}}} = \dfrac{M_{\mathrm{a}}}{M_{\mathrm{d}}}\dfrac{\tau_{\mathrm{KH,d}}}{\beta}\quad.
\end{equation}
Since the global thermal timescale scales as $\tau_{\mathrm{KH}}\propto M^{-3/2}$ all systems with $q_{\mathrm{i}} < 1$ will have accretors that are out of thermal equilibrium (at least at the onset of mass transfer) when mass transfer is fully conservative ($\beta = 1$). To find the critical mass ratios below which this is still the case when mass transfer is non-conservative ($\beta < 1$), we look for systems for which $\tau_{\mathrm{accr}} < \tau_{\mathrm{KH,a}}$. Using the expressions above, we find that $q_{\mathrm{i}}^{\mathrm{crit}} = \beta^{2/5}$. In the case of our model, this means that we expect to find similar results for $q_{\mathrm{i}} < q_{\mathrm{i}}^{\mathrm{crit}} = 0.5^{2/5} \approx 0.76$. Hence, we can estimate that systems with higher initial mass ratios will have accretors that can in principle retain thermal equilibrium and have BV frequency profiles more akin to those shown in Fig.~\ref{fig:combined_peak}.

\subsection{The decrease in outer BV frequency peak magnitude with increasing semiconvective efficiency}\label{sec:effect_on_peak}

With the exact origin of the accretion-induced outer BV frequency peak established, we are now well-equipped to explain the decrease in magnitude of this peak with increasing $\alpha_{\mathrm{sc}}$, which was previously discussed in Sect.~\ref{sec:asteroseismic_predicitions}. Comparing the $\alpha_{\mathrm{sc}}=0.1$ and $\alpha_{\mathrm{sc}}=10$ models in Fig.~\ref{fig:all_kipps_asc}, we see that they reach comparable convective core masses during rejuvenation. This means that the steep composition gradient responsible for the outer peak is located at similar mass coordinates in both models. The difference between the models is that in the $\alpha_{\mathrm{sc}}=10$ model, the upper part of this composition gradient is smoothed out by the 100 times as efficient semiconvective mixing. As a result, only the lower half (in terms of mass coordinate) of the steep composition gradient between the newly formed convective core and envelope is left intact. This causes a smaller, narrower peak in the BV frequency profile of the $\alpha_{\mathrm{sc}}=10$ model. The upper half of the original steep chemical composition gradient becomes less steep because of efficient semiconvective mixing. This upper half of the chemical composition gradient gives rise to an additional, smaller BV frequency peak around $u=0.3$ ($m/\msun=0.92$), as can be seen in Fig.~\ref{fig:BV_and_FT_Xc30}.

Between $\alpha_{\mathrm{sc}}=0.0\text{--}0.1$, the opposite trend occurs, namely, the maximum value of the outer BV frequency peak increases with $\alpha_{\mathrm{sc}}$. At $X_{\mathrm{c}}=0.30$, this value increases from $54$~d$^{-1}$ to $64$~d$^{-1}$ between $\alpha_{\mathrm{sc}}=0.0$ and $\alpha_{\mathrm{sc}}=0.1$, while there is a negligible change in this value between $\alpha_{\mathrm{sc}}=0.01$ and $\alpha_{\mathrm{sc}}=0.1$ (the former is not shown in Fig.~\ref{fig:BV_and_FT_Xc30}). We attribute this to minor differences in the (semi-)convective region above the convective cores of these models.

\subsection{Fourier transforms of PSPs}\label{sec:fourier}

In addition to the BV frequency profiles, the panels in Fig.~\ref{fig:BV_and_FT_Xc30} contain a second y-axis showing the Fourier transform (FT) of the normalised PSP as a function of the radial order $n$, which is
\begin{equation}
   \Delta P_{\ell=1}^{\mathrm{norm}}(n) = \frac{\Delta P_{\ell=1}(n) - \Pi_{\ell=1}}{\Pi_{\ell=1}}\,,
\end{equation}
for each model. We used the \texttt{numpy} \citep{Harris2020} fast Fourier transform algorithm, \texttt{fft}, applied to $\Delta P_{\ell=1}^{\mathrm{norm}}(n)$. As demonstrated in \citet{Guo2026}, this method of taking the FT of a PSP allows one to characterise the (relative) amplitude and width of variability features in radial order space because $\Delta n=1/u(r)$. In essence, $\Delta n$ acts as the `frequency' of the PSP variability in this framework \citep{Guo2026}. The proof-of-concept work by \citet{GuoAerts2025} shows how this FT method can be used to constrain core-hydrogen mass fractions in stars with observed PSPs. Moreover, as demonstrated in \citet{Wu2026} for models of accretors, this FT helps to quantitatively identify additional components in the PSP variability, which show up as additional peaks in the FT. 

Entirely consistent with aforementioned works, we see in Fig.~\ref{fig:BV_and_FT_Xc30} and the equivalent figures in Appendix~\ref{app:other_seismic_figures} that the locations of the FT peaks on the $u(r)$-axis correspond to the locations of sharp variations in $\tilde{N}$, which are responsible for mode trapping, and hence, the PSP variability. In general, mode trapping occurs when $\tilde{N}$ varies strongly over a radial range that is smaller or comparable to the local wavelength of the gravity modes \citep{Cunha2020}. This trapping alters the mode's properties, such as its period. In line with \citet{Wu2026}, we see that whereas the single stars typically have one well-defined peak in their FT at the interface between the near-core chemical gradient and the stellar envelope (at $u \approx 0.3$ in Fig.~\ref{fig:BV_and_FT_Xc30}), the accretor models show multiple FT peaks because of the multiple BV frequency peaks discussed above. These multiple FT peaks are a clear sign of the multiple components of PSP variability for accretor stars, as described qualitatively in Sect.~\ref{sec:asteroseismic_predicitions}. The inner peak (at lower $u$) in the FT of the $\alpha_{\mathrm{sc}}=0.1$ model's PSP is caused by the transition between the inner and outer BV frequency peaks. Based on our explanation for the exact shape of this double-peaked BV frequency feature in Sect.~\ref{sec:peak_origins}, the magnitude of this peak in the FT in general depends critically on how far the accretor is out of thermal equilibrium during accretion.

The relative heights of the FT peaks tell us something about the relative amplitudes of the PSP variation caused by these features in the BV frequency. The location of the peaks in $u(r)$-space hold information about the width $\Delta n$ of the different variability features, with smaller $\Delta n$ corresponding to PSPs that vary more rapidly in terms of $n$. From the definition of $\Delta n$ above, we see that FT peaks at values of $u(r)$ closer to 0.5 correspond to faster variations in the PSP. In the limit of $u=0.5$, the width of the variation in radial order space becomes $\Delta n = 2$, which, given that $n$ only takes integer values, corresponds to the Nyquist frequency of our PSPs. FT features appearing at $u>0.5$, as seen at later evolutionary stages (see Figs.~\ref{fig:BV_and_FT_Xc47}--\ref{fig:BV_and_FT_Xc1}), are reflected into the $u<0.5$ region \citep{Montgomery2003}.

Despite the fact that the PSPs shown in the previous section do not show clear qualitative variations with $\alpha_{\mathrm{sc}}$, their FTs reveal this variation in a quantitative manner. We see that the relative amplitudes of the inner and outer FT peaks change with $\alpha_{\mathrm{sc}}$, and that the additional FT peaks at higher $u(r)$ gain in relative magnitude, even overtaking the other peaks in the $\alpha_{\mathrm{sc}}=10$ model. Right after accretion, at $X_{\mathrm{c}}=0.47$ (Fig.~\ref{fig:BV_and_FT_Xc47}), the FT peaks originating from the short-lived envelope convection zones during accretion are more dominant since they have not yet been mixed out by the imposed envelope mixing. At later main-sequence ages between $X_{\mathrm{c}}=0.10$ and $X_{\mathrm{c}}=0.01$ (Figs.~\ref{fig:BV_and_FT_Xc10} and \ref{fig:BV_and_FT_Xc1}), the near core region with the strong variations in $\tilde{N}$ lie closer to or beyond $u(r) = 0.5$. In the latter case, we do not expect to be able to sample such features, given that $n$ is sampled in integers.

\section{Isolated effect of semiconvection: models without convective boundary mixing}\label{sec:results_noos}

\begin{table}
\centering
\caption{Helium (He) core masses at TAMS of the accretor and single-star models without overshooting}
\label{tab:He_core_masses_noos}
\begin{tabular}{lc@{\hspace{5em}}lc}
\hline\hline
$\alpha_{\mathrm{sc}}$ & $M_{\mathrm{He}}^{\mathrm{TAMS}}$ & $M_{\mathrm{single}}$  & $M_{\mathrm{He}}^{\mathrm{TAMS}}$ \\
                       & $[\mathrm{M}_{\odot}]$            & $[\mathrm{M}_{\odot}]$ & $[\mathrm{M}_{\odot}]$            \\ \hline
0.0                    & 0.319                             & 3.45                   & 0.325                             \\
0.001                  & 0.320                             & 3.46                   & 0.326                             \\
0.01                   & 0.321                             & 3.47                   & 0.327                             \\
0.1                    & 0.328                             & 3.48                   & 0.328                             \\
1                      & 0.330                             & 3.49                   & 0.329                             \\
10                     & 0.333                             & 3.50                   & 0.330                             \\
100                    & 0.336                             & 3.51                   & 0.332                             \\
1000                   & 0.337                             & 3.52                   & 0.333                             \\
Schw.                  & 0.339                             & 3.53                   & 0.334                               \\
                       &                                   & 3.54                   & 0.335                               \\
                       &                                   & 3.55                   & 0.337                             \\
\hline\hline
\end{tabular}
\end{table}

\begin{figure*}
\begin{centering}
\includegraphics[width=0.95\textwidth]{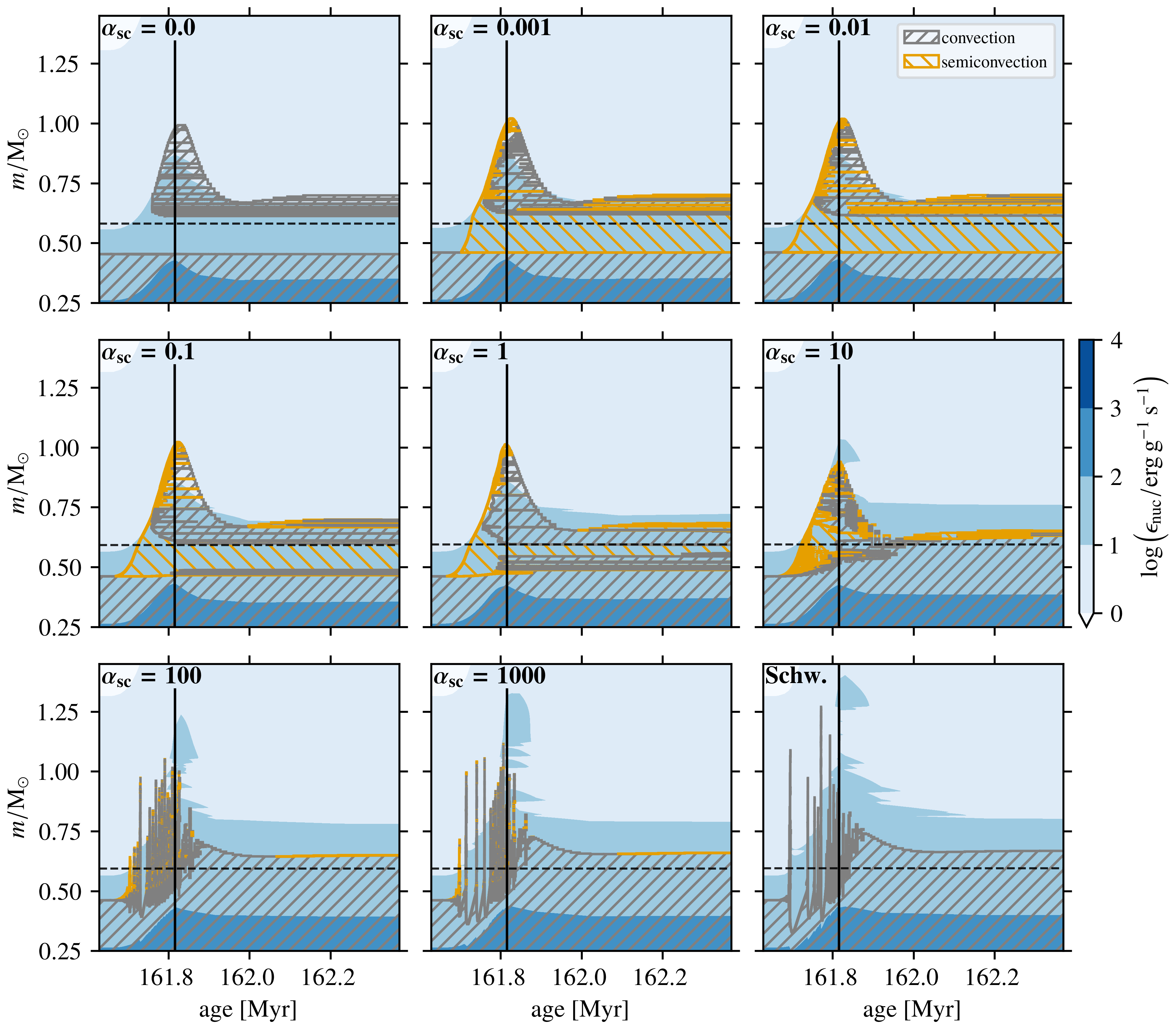}
\par\end{centering}
    \caption{Same as Fig.~\ref{fig:all_kipps_asc} but for the models without CBM. The evolution of the models is shown up to two global thermal timescales after the end of mass transfer, instead of up to one time this timescale as in Fig.~\ref{fig:all_kipps_asc}.}
    \label{fig:all_kipps_noos}
\end{figure*}

\begin{figure}
 \includegraphics[width=\columnwidth]{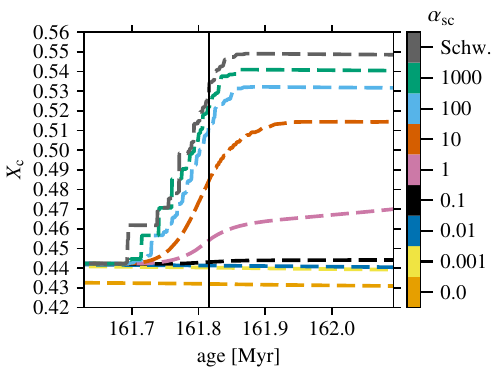}
 \caption{Same as Fig.~\ref{fig:center_h1_asc} but for the models without CBM.}
 \label{fig:center_h1_noos}
\end{figure}

\begin{figure}
 \includegraphics[width=0.95\columnwidth]{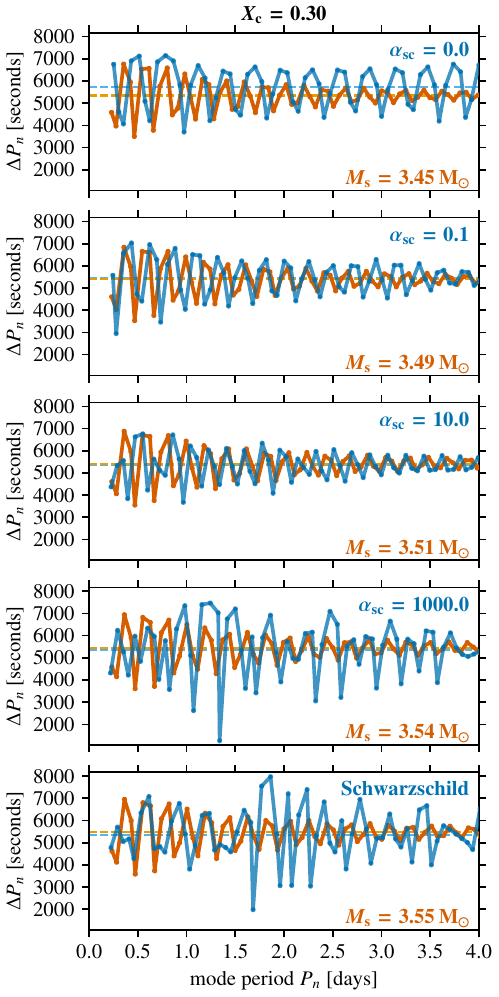}
 \caption{Same as Fig.~\ref{fig:PSP_asc_Xc30} but for the models without overshooting.}
 \label{fig:PSP_noos_Xc30}
\end{figure}

\begin{figure}
 \includegraphics[width=0.95\columnwidth]{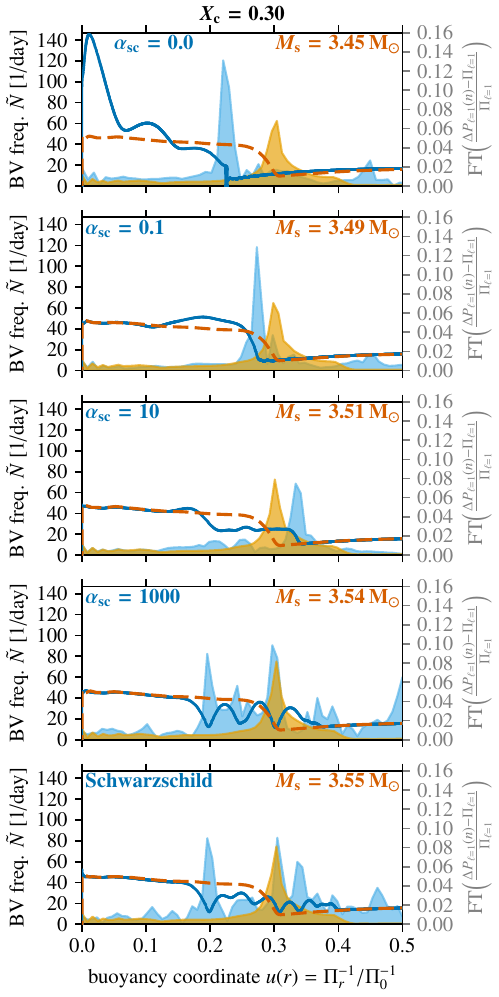}
 \caption{Same as Fig.~\ref{fig:BV_and_FT_Xc30} but for models without overshooting.}
 \label{fig:BV_and_FT_Xc30_noos}
\end{figure}

In this section, we repeat the analysis from Sect.~\ref{sec:results_standard} for accretor and equivalent single-star models without overshooting. Contrary to what we found for the original models, we now see the classical story from \citet{Braun1995} emerge from the Kippenhahn diagrams of our models, which are shown in Fig.~\ref{fig:all_kipps_noos}. In models with no or low-efficiency semiconvective mixing, we find that the convective core fails to grow due to the mean molecular weight gradient on top of it. Simultaneously, a convective shell appears above the mass coordinate of the largest extent of the pre-mass-transfer convective core, which is $m/\msun=0.593$. The models with $\alpha_{\mathrm{sc}}=0.0$, $0.001$, and $0.01$ show no signs of rejuvenation, due to their inefficient semiconvective mixing. This is further supported by the fact that their central hydrogen mass fraction stays virtually constant (Fig.~\ref{fig:center_h1_noos}) and from their He core masses at TAMS (Table~\ref{tab:He_core_masses_noos}). The He cores of these low $\alpha_{\mathrm{sc}}$ models are under-massive for their total mass. Starting from the $\alpha_{\mathrm{sc}}=0.1$ models, rejuvenation does take place with increasing efficiency, which is reflected by the He core masses at TAMS. When $\alpha_{\mathrm{sc}} \geq 1$ the accretors have over-massive cores. 

We find that in all models that rejuvenate, a considerable part of the rejuvenation occurs after mass transfer ends. The speed at which this rejuvenation takes place is set by the semiconvective efficiency, as is apparent from the Kippenhahn diagrams in Fig.~\ref{fig:all_kipps_noos}. The convective shells remain present for more than two thermal timescales after the end of mass transfer in the models with $\alpha_{\mathrm{sc}} \in [0.1, 1, 10]$. They disappear before the core has grown to its largest post-mass-transfer extent and then starts receding again. In the models that do not rejuvenate, the convective shells persist until the core recedes again on the continuation of the MS. Analogous to the original models with overshooting in Sect.~\ref{sec:results_noos}, we find that models with $\alpha_{\mathrm{sc}} \geq 100$ rejuvenate in bursts and have multiple, short-lived convective shells that are able to extend into the region that (previously) had a chemical composition gradient. The mechanism here is most likely that of the secular heat-engine found for the original models, though it seems to operate on a slightly different timescale. Lastly, we find that overall, the extent of rejuvenation is smaller without overshooting than with overshooting. This again highlights the non-negligible role of overshooting in the rejuvenation process.

\subsection{Asteroseismic predictions and PSP Fourier transforms}
Looking at the BV frequency profiles in Fig.~\ref{fig:BV_and_FT_Xc30_noos}, the first striking difference with the original models is that the typical double-peaked feature associated with accretion and rejuvenation is missing in all models. We discuss these model below.

The model with $\alpha_{\mathrm{sc}}=0.0$, which does not rejuvenate, has a $\tilde{N}$-profile that deviates strongly from that of its equivalent single star. During most of the accretion phase, this model's $\mu$-profile remains unaffected. Only when the convective shell appears, a somewhat sharp feature appears in the $\mu$-profile at the bottom of this zone. This feature gives rise to the relatively strong variation in $\tilde{N}$ around $u\approx 0.22$, which can be seen in Fig.~\ref{fig:BV_and_FT_Xc30_noos}. After the accretion ends and the accretor regains thermal equilibrium, the convective core recedes again. This lead to a $\mu$-gradient and, hence, the strong peak in $\tilde{N}$ around $u\approx0.01$. The lower peak in $\tilde{N}$ around $u\approx0.1$ is a remnant of the original $\mu$-gradient from the pre-mass-transfer receding convective core. It is the variation in $\tilde{N}$ at $u\approx 0.22$ that is responsible for mode trapping and hence PSP variation in this model. This can be appreciated in Fig.~\ref{fig:BV_and_FT_Xc30_noos} from the FT of the associated PSP, which is shown in Fig.~\ref{fig:PSP_noos_Xc30}. Even though this $\tilde{N}$ variation get smoothed out by the imposed envelope mixing at later stages of the accretor's MS evolution (see Figs.~\ref{fig:PSP_noos_Xc42}--\ref{fig:PSP_noos_Xc1}), it remains responsible for mode trapping. Due to the strongly deviating $\tilde{N}$-profile of the $\alpha_{\mathrm{sc}}=0.0$ model compared to that of its equivalent single star, its $\Pi_{\ell=1}$ value deviates considerably ($\sim 500\,\mathrm{s}$) from that of the equivalent single star, as seen in Fig.~\ref{fig:PSP_noos_Xc30}. This difference is smaller at other points during the MS, as shown in Figs.~\ref{fig:PSP_noos_Xc42}--\ref{fig:PSP_noos_Xc1}.

As described earlier in this section, the model with $\alpha_{\mathrm{sc}}=0.1$ rejuvenates (albeit not fully) several thermal timescales after the end of mass transfer. At this point, the convective core has expanded up to the bottom of the convective shell, which disappears once the core starts receding again. Similar to the $\alpha_{\mathrm{sc}}=0.0$ model, the feature in the star's $\tilde{N}$-profile responsible for mode trapping is a remnant of the connection between the convective shell and the radiative region below. As in the $\alpha_{\mathrm{sc}}=0.0$ model, this connection causes a somewhat sharp feature in the $\mu$-profile and a strong variation in the $\tilde{N}$-profile. The highest peak in the $\tilde{N}$-profile around $u\approx 0.2$, stretched out because of the choice of coordinate system, is not caused by the recession of the convective core on the MS, but is also a feature resulting from $\mu$-gradient left behind by the shrinking convective shell. The part of the $\tilde{N}$-profile to the left of this peak, which is at $u \lesssim 0.11$, closely follows that of the equivalent single star. This part is caused by the receding convective core on the MS after mass transfer.

The $\alpha_{\mathrm{sc}}=10$ model has yet another shape in its $\tilde{N}$-profile compared to the two models discussed above. The strong variation responsible for mode trapping, found at $u\approx 0.34$ is again caused by the convective shell. Its location corresponds roughly to that of the maximum extent of the shell. At $0.2 \lesssim u \lesssim 0.34$, we find an almost flat $\tilde{N}$-profile caused by the convective shell. We note that this part of the $\tilde{N}$-profile is smoothed out progressively by the imposed envelope mixing. The largest peak in $\tilde{N}$ is found at the location where the bottom of the convective shell used to be and the convective core expanded to before receding again. Just as for the $\alpha_{\mathrm{sc}}=0.1$ model, the left-most part of the $\tilde{N}$-profile closely follows that of the accretor's equivalent single star.

What follows from the description of the models with $\alpha_{\mathrm{sc}}=0.0\text{--}10$ above, is that they all only have one location in the star that is responsible for mode trapping. This is especially clear from the FT of the PSPs shown in Fig.~\ref{fig:BV_and_FT_Xc30_noos} for $X_{\mathrm{c}}=0.30$, and the equivalent Figs.~\ref{fig:BV_and_FT_Xc42_noos}, ~\ref{fig:BV_and_FT_Xc10_noos}, and ~\ref{fig:BV_and_FT_Xc1_noos} for $X_{\mathrm{c}}=0.42$, $0.10$, and $0.01$, respectively. The location of the dominant peak in the FTs is set by the location of the bottom of the convective shell that appears during accretion, and is therefore at the same mass coordinate in all models. This is because the location of the bottom of the convective shell is roughly set by the largest extent of the pre-mass-transfer convective core. Despite being at the same mass coordinate, we find the location of the dominant FT peak at different $u$-coordinates for the different models, by virtue of its definition and the variety in $\tilde{N}$-profiles across the models. For example, for the $\alpha_{\mathrm{sc}}=0.0$ and $0.1$ models, the FT peak is located at a lower $u$ than for their respective equivalent single stars. This corresponds to the somewhat larger width of the variation of the accretors' PSPs. Whereas, the $\alpha_{\mathrm{sc}}=10$ model has a FT peak at a $u$-coordinate that is higher than that of its equivalent single star's FT peak, which corresponds to more rapid variation in the former's PSP. Because of the proximity in terms of $u$ for the mode trapping location in the accretors and their equivalent single stars, the resulting PSPs are less distinguishable from each other than they are for the original models that include overshooting discussed in Sect.~\ref{sec:asteroseismic_predicitions}. Nevertheless, the point that the PSP variation has larger amplitudes for accretors than for single stars, made before by \citet{Wagg2024}, \citet{Miszuda2025betaCep}, and \citet{Wu2026}, seems to hold, even without the inclusion of overshooting for different semiconvection efficiencies.

This brings us to an important point. Let us assume that in real stars, rejuvenation occurs in a way that more closely resembles that found for the models without overshooting. In other words, let us assume that in reality, CBM does not enable convective core growth during accretion and that this is solely enabled by (moderately efficient) semiconvective mixing as in the picture of \citet{Braun1995}. In that case, based on the present models, we only expect to find one dominant peak in the FT of an accretor's PSP. Contrary to what was found based on our original models and the models from \citet{Wu2026}, this would mean that the appearance of multiple dominant FT peaks to identify accretors loses its diagnostic power. Therefore, the direct observational test predicted from our models on the importance of CBM (and minor role of $\alpha_{\mathrm{sc}}$) in accretor seismology is the presence of multiple peaks in the FT of the PSP.

We now focus on the model with $\alpha_{\mathrm{sc}} = 1000$ and the model that uses the Schwarzschild condition for convection. Due to the more complex evolution of the near-core region during and post accretion in these models, notably because of the appearance of multiple short-lived convective shells, the $\tilde{N}$-profiles contain multiple peaks. Nevertheless, just as their counterparts with less efficient semiconvective mixing, their $\tilde{N}$-profiles miss the typical double-peak structure caused by CBM-enabled rejuvenation found in our original models. We identify two dominant peaks in the FTs in Fig.~\ref{fig:BV_and_FT_Xc30_noos} that point to two dominant mode-trapping locations. The inner peak, at $u\approx 0.2$ in both models, originates from the receding convective core. The outer one at $u\approx 0.3$ originates from the convective shells. The two-component PSPs variability implied by these two dominant FT peaks can be observed in Fig.~\ref{fig:PSP_noos_Xc30}.

Lastly, it is worth noting that the sharp features in $\mu$ at the bottom of the convective shells ultimately responsible for mode trapping in the $\alpha_{\mathrm{sc}}=0.0\text{--}10$ models might just be the result of a lack of CBM included for such convective regions. Our current models only model CBM for convective core regions. One can imagine that, for example, even a small amount of exponential overshooting at the bottom of this convective zone could soften this feature in the $\mu$-profile and, hence, a weaker variation in the $\tilde{N}$-profile. In other words, the mode trapping might be caused and/or amplified by the lack of CBM included in our models for convective shells.

\section{Summary \& conclusions}\label{sec:conclusions}

In this work, we set out to find the effect of the efficiency of semiconvective mixing on the asteroseismic imprints of accretion in intermediate and high-mass stars previously described in the literature. Despite the relatively straightforward setup of this parameter study, we arrive at the following conclusions, which paint a somewhat more complex picture.\\

In accretor models using the Ledoux criterion for convection, convective boundary mixing in the form of overshooting is the main enabler of rejuvenation and causes all models to rejuvenate. This is a departure from the classical picture from \citet{Braun1995}, which does not consider CBM and in which the semiconvective efficiency largely determines the occurrence and amount of rejuvenation. In our models without CBM, the occurrence and amount of rejuvenation are dictated by the efficiency of semiconvection, following \citet{Braun1995}. This strong influence of CBM highlights the necessity for dedicated follow-up work focussed on varying amounts and prescriptions for CBM. 

The reality is that step and exponential overshooting, both independent of the near-core chemical composition profile by virtue of their implementation, are the current standard in stellar structure and evolution computations. As a result, we expect all accretors in such computations to rejuvenate at least to some extent. It is essential to keep this in mind while interpreting the results of accretor evolution computations, given that it has yet to be established if and to what extent rejuvenation occurs in all types of mass-transferring binary systems.

The effects of other types of mixing, such as rotation-induced mixing, are also expected to influence these results (Miszuda et al. in prep.). We can already see what the effect of extra mixing could be from the parameter study in \citet[][Appendix A]{Wagg2024}, in which they varied the minimum envelope mixing, with a diffusive mixing coefficient of $D_\mathrm{min}\sim 100\,\mathrm{cm}^{2}\mathrm{s}^{-1}$ altering the shape of the outer BV frequency peak and reducing the amplitude of the second PSP variability component. However, for a different binary system and a $D_\mathrm{min} = 100\,\mathrm{cm}^{2}\mathrm{s}^{-1}$, \citet{Miszuda2025betaCep} do not find a strong flattening of the outer peak in the BV frequency profile of their accretor model. This only enforces our conclusion that systematic studies of the effect of mixing prescriptions and efficiencies are imperative. 

Furthermore, we stress that it is unlikely that CBM acting in real stars would be unaffected by the chemical stratification of the near-core region, as is assumed in the overshooting prescriptions used in our (and most other) models. The impact of this chemical stratification is to presumably dampen the overshooting material and reduce the size of the CBM region. In that sense, the CBM-driven rejuvenation reported here could be a consequence of the way in which near-core mixing is implemented in stellar structure and evolution codes. Nevertheless, any code or prescription that includes some kind of sufficiently efficient CBM will diminish the chemical composition gradient outside the core to some extent and allow the convective core to grow into chemically homogeneous regions. In which situations (e.g. evolutionary stages) CBM is sufficient to facilitate convective core growth during accretion will ultimately have to be answered by dedicated 3D simulations \citep[see e.g.][]{Scott2021, Morison2024}, which is beyond the scope of this work.

Another aspect of accretion that influences the occurrence and amount of rejuvenation is the accreted mass relative to the initial mass \citep{Braun1995}. This ratio of accreted over initial mass is fixed in all of our models to $0.5/3.0 \approx 0.17$, but its effect is explored in detail in \citet{Schneider2024}. Ultimately, the effect of the relative accreted mass on rejuvenation should be studied in combination with that of semiconvection and CBM to arrive at a more complete picture of the structural and seismic imprints of accretion.\\

The final double-peaked Brunt-V\"ais\"al\"a $\tilde{N}$ profile of accretors with CBM, which has been previously identified by \citet{Wagg2024} and \citet{Miszuda2025betaCep}, is not only set by the overall convective core growth, but is strongly influenced by the transient behaviour of the convective core during thermal relaxation after the end of mass transfer. In our work, we find that accretors which maintain thermal equilibrium do not contract considerably after mass transfer ceases and have considerably different $\tilde{N}$-profiles, akin to those in \citet{Wu2026}. Moreover, the double-peaked shape of the $\tilde{N}$-profile is absent in accretor models without CBM.\\

Irrespective of whether CBM is included or not, models with semiconvective efficiencies of $\alpha_{\mathrm{sc}} \geq 100$ (including the models using the Schwarzschild criterion for convection) have strong, burst-like rejuvenation episodes. These rejuvenation bursts are a consequence of step-like mean molecular weight profiles caused by the contrast between negligible and efficient mixing in neighbouring layers. These steps in mean molecular weight lead to relatively large jumps in the opacity and, hence, the efficient blocking of radiative flux. The resulting secular heat-engine-like mechanism causes efficient rejuvenation in bursts thanks to extended, short-lived convection zones. We speculate that in these models, the effect of semiconvection is indirect, but has a stronger influence on the amount of rejuvenation than in the lower-efficiency models without bursts. It remains to be confirmed whether this secular heat-engine mechanism for rejuvenation in bursts is genuinely physical, or a result of the sharp transition in diffusive mixing coefficients between neighbouring layers in evolution codes such as \texttt{MESA}. \\

For all accretors, with and without CBM, the amplitude of the period spacing pattern variations is larger than their respective equivalent single stars. PSPs of accretors with CBM show a clear second component in their variability. However, with typical uncertainties on mode periods and period spacings in mind, it is virtually impossible to identify clear qualitative differences between the PSPs of models with different semiconvective efficiencies. These differences with $\alpha_{\mathrm{sc}}$ can be made quantitative by looking at the Fourier transforms of the PSPs. These FTs clearly quantify the differences in the PSPs caused by varying semiconvection and the presence of CBM.
We reaffirm the conclusion of \citet{Wu2026} that FTs show great potential for disentangling the multiple components in the PSP variability of accretors and confirm their status as post-accretion stars.\\

Overall, our work demonstrates how sensitive asteroseismic predictions are to the assumptions made for the underlying stellar structure and evolution models, but specifically in the context accretor stars in this work. Although we show that the asteroseismic imprint of accretion first reported in \citet{Wagg2024}, \citet{Miszuda2025betaCep}, and \citet{Wu2026} is relatively robust against the efficiency of semiconvection, the effect of CBM ought to be explored in more detail. On top of that, we demonstrate how sensitive the internal structure and, hence, asteroseismic properties, of accretors can be to the kind of mass transfer (rapid or slow) and the post-mass transfer thermal relaxation. Given the large variety of mass-transferring binary systems expected in nature \citep[see e.g.][and references therein]{Marchant2024a, Schneider2025review}, a more general systematic exploration of the seismic properties of accretors is needed to arrive at a unified picture of the asteroseismic imprints of accretion. 

\section*{Acknowledgements}
The authors thank the anonymous referee for the constructive and encouraging feedback, which helped us improve this manuscript.
The authors thank Norbert Langer for the fruitful discussion and his useful feedback.
The authors gratefully acknowledge UK Research and Innovation (UKRI) in the form of a Frontier Research grant under the UK government's ERC Horizon Europe funding guarantee (SYMPHONY; PI Bowman; grant number: EP/Y031059/1), and a Royal Society University Research Fellowship (PI Bowman; grant number: URF{\textbackslash}R1{\textbackslash}231631).We used the following software for the analysis in this work: \texttt{PyGYRE} \citep{Townsend2020pygyre}, \texttt{PyMesaReader} \citep{PyMesaReader2017}, \texttt{MPI for Python} \citep{Dalcin2005,Dalcin2021}, \texttt{Astropy} \citep{Astropy2013,Astropy2018,Astropy2022}, \texttt{NumPy} \citep{Harris2020}, \texttt{SciPy} \citep{Scipy2020}, and \texttt{Plotly} \citep{Plotly}. We used \texttt{Matplotlib} \citep{Hunter2007} for creating the plots in this manuscript. We have made extensive use of the Modules for Experiments in Stellar Astrophysics \citep[\texttt{MESA},][\url{https://mesastar.org/}]{Paxton2011, Paxton2013, Paxton2015, Paxton2018, Paxton2019, Jermyn2023} and \texttt{GYRE} \citep[][\url{https://gyre.readthedocs.io/en/stable/}]{Townsend2013, Townsend2018} codes in this work. 

\section*{Data Availability}
Data products that support the results in this paper are publicly available via the Zenodo repository: \url{https://zenodo.org/records/20623400} (\url{https://doi.org/10.5281/zenodo.20623400}).

\bibliographystyle{mnras}
\bibliography{references}

\newpage
\FloatBarrier

\appendix

\section{Period spacing patterns at different ages}\label{app:other_psps}

\begin{figure}
 \includegraphics[width=0.95\columnwidth]{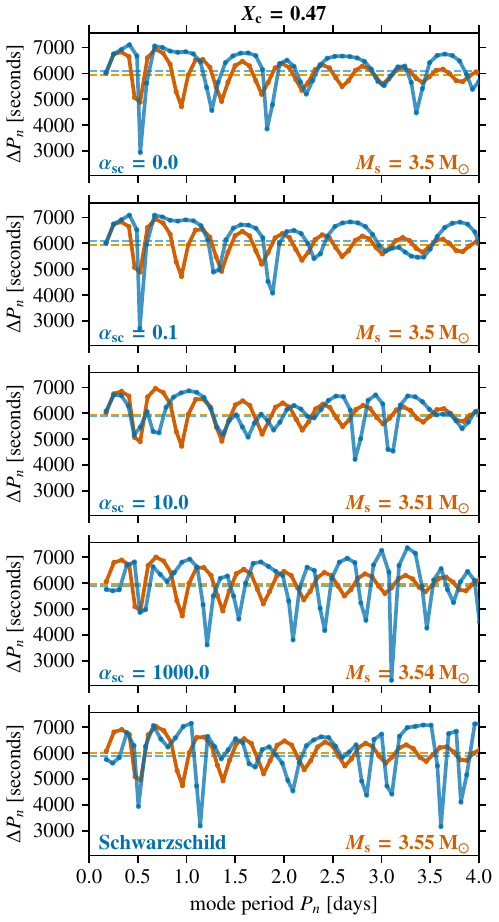}
 \caption{Same as Fig.~\ref{fig:PSP_asc_Xc30} but for $X_{\mathrm{c}} = 0.47$.}
 \label{fig:PSP_asc_Xc47}
\end{figure}

\begin{figure}
 \includegraphics[width=0.95\columnwidth]{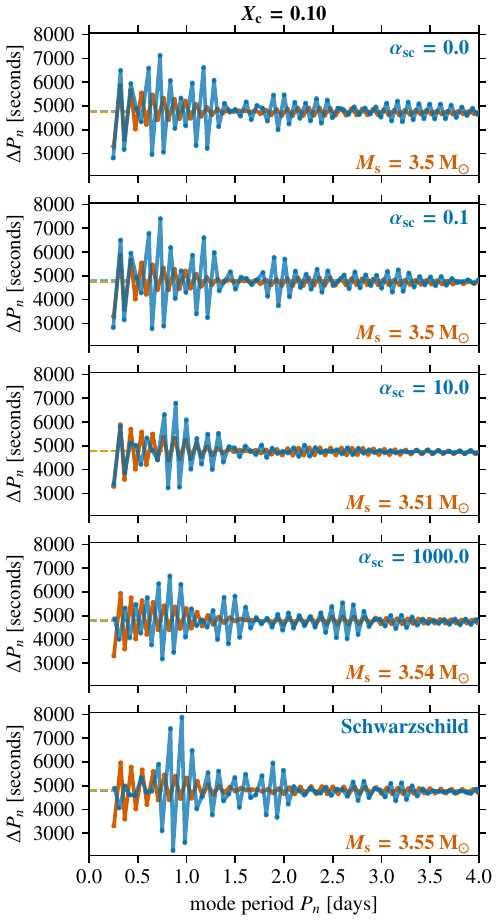}
 \caption{Same as Fig.~\ref{fig:PSP_asc_Xc30} but for $X_{\mathrm{c}} = 0.10$.}
 \label{fig:PSP_asc_Xc10}
\end{figure}

\begin{figure}
 \includegraphics[width=0.95\columnwidth]{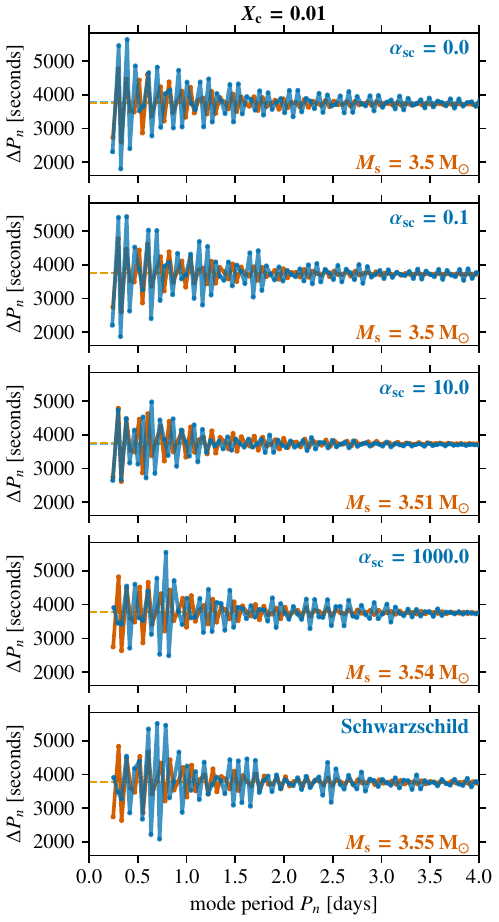}
 \caption{Same as Fig.~\ref{fig:PSP_asc_Xc30} but for $X_{\mathrm{c}} = 0.01$.}
 \label{fig:PSP_asc_Xc1}
\end{figure}

\begin{figure}
 \includegraphics[width=0.95\columnwidth]{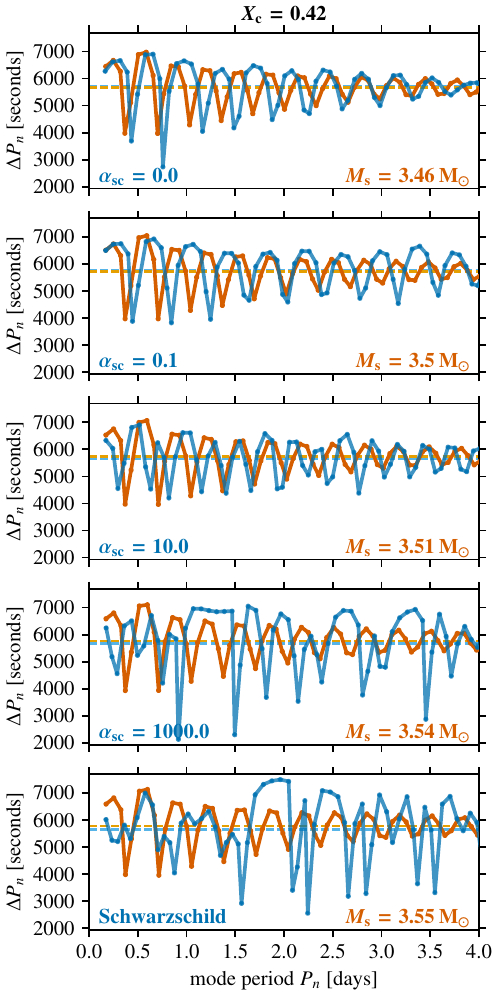}
 \caption{Same as Fig.~\ref{fig:PSP_noos_Xc30} but for $X_{\mathrm{c}} = 0.42$.}
 \label{fig:PSP_noos_Xc42}
\end{figure}

\begin{figure}
 \includegraphics[width=0.95\columnwidth]{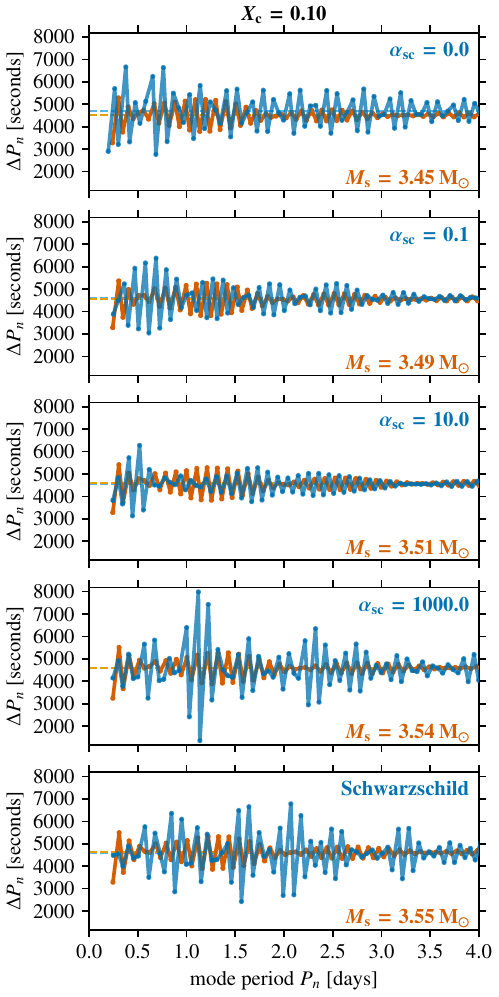}
 \caption{Same as Fig.~\ref{fig:PSP_noos_Xc30} but for $X_{\mathrm{c}} = 0.10$.}
 \label{fig:PSP_noos_Xc10}
\end{figure}

\begin{figure}
 \includegraphics[width=0.95\columnwidth]{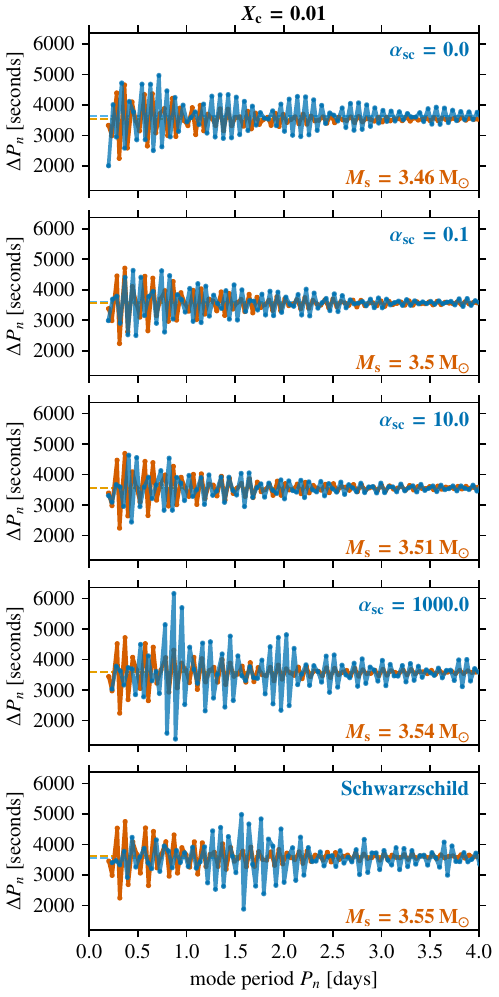}
 \caption{Same as Fig.~\ref{fig:PSP_noos_Xc30} but for $X_{\mathrm{c}} = 0.01$.}
 \label{fig:PSP_noos_Xc1}
\end{figure}

\FloatBarrier

\section{BV frequency profiles and PSP FTs at different ages}\label{app:other_seismic_figures}

\begin{figure}
 \includegraphics[width=0.95\columnwidth]{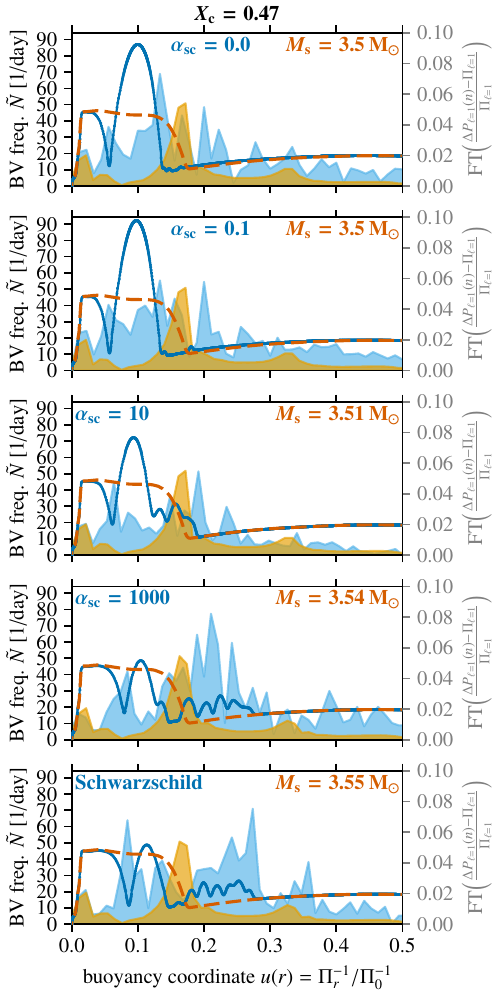}
 \caption{Same as Fig.~\ref{fig:BV_and_FT_Xc30} but for $X_{\mathrm{c}} = 0.47$.}
 \label{fig:BV_and_FT_Xc47}
\end{figure}

\begin{figure}
 \includegraphics[width=0.95\columnwidth]{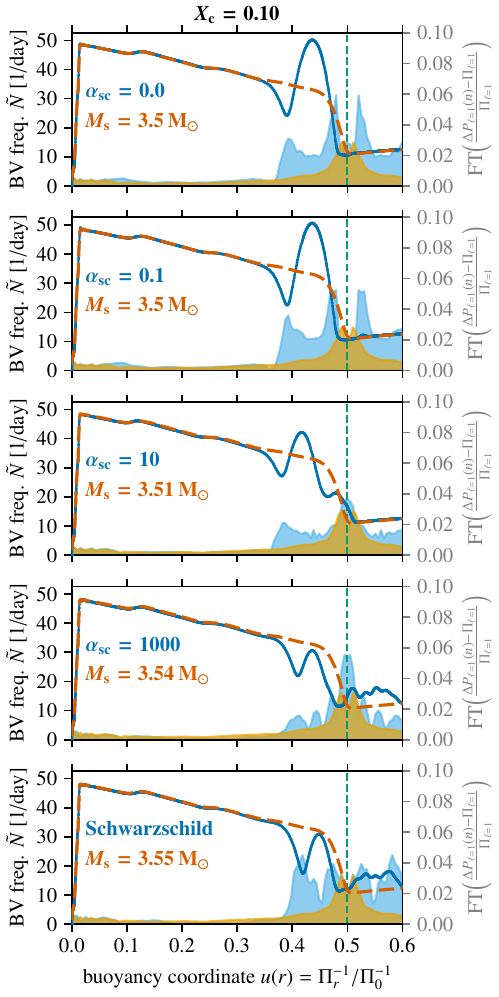}
 \caption{Same as Fig.~\ref{fig:BV_and_FT_Xc30} but for $X_{\mathrm{c}}=0.10$. The vertical dashed green line indicates the Nyquist frequency at $u(r) = 0.5$.}
 \label{fig:BV_and_FT_Xc10}
\end{figure}

\begin{figure}
 \includegraphics[width=0.95\columnwidth]{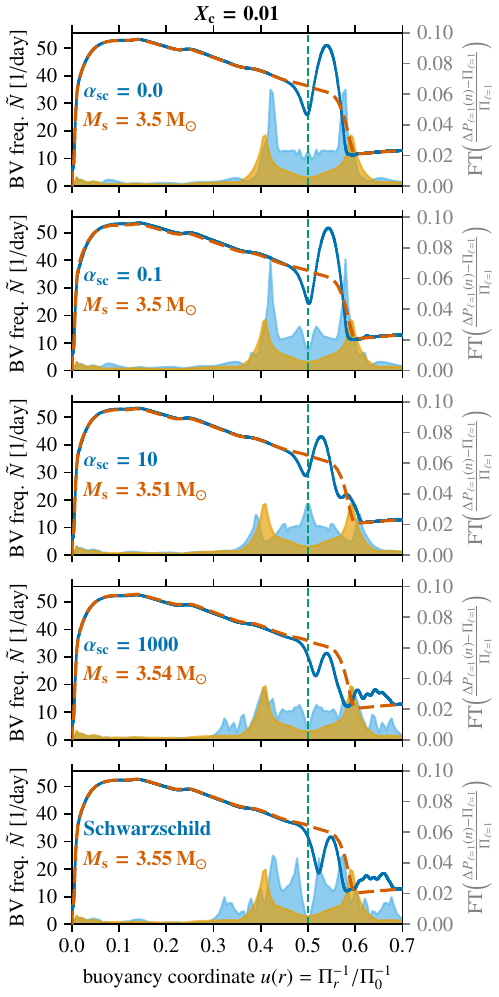}
 \caption{Same as Fig.~\ref{fig:BV_and_FT_Xc30} but for $X_{\mathrm{c}}=0.01$. The vertical dashed green line indicates the Nyquist frequency at $u(r) = 0.5$.}
 \label{fig:BV_and_FT_Xc1}
\end{figure}

\begin{figure}
 \includegraphics[width=0.95\columnwidth]{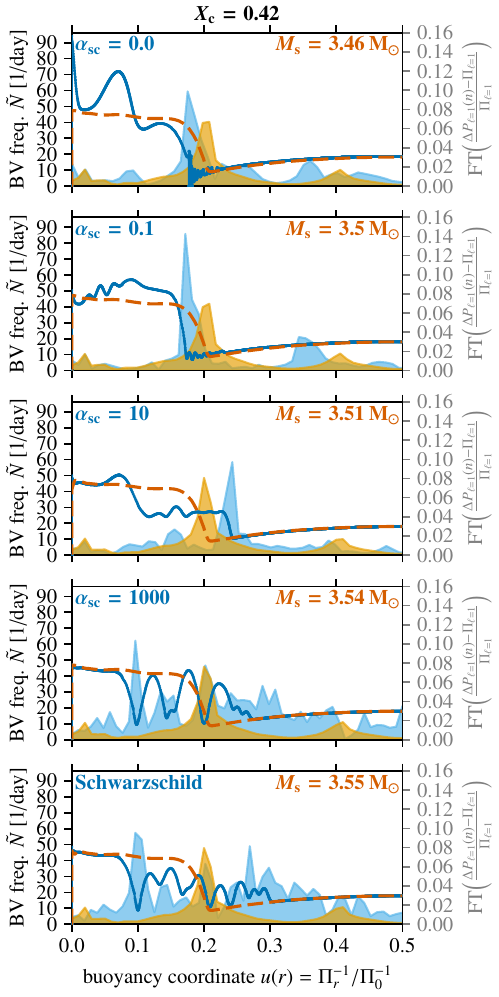}
 \caption{Same as Fig.~\ref{fig:BV_and_FT_Xc30_noos} but for $X_{\mathrm{c}} = 0.42$.}
 \label{fig:BV_and_FT_Xc42_noos}
\end{figure}

\begin{figure}
 \includegraphics[width=0.95\columnwidth]{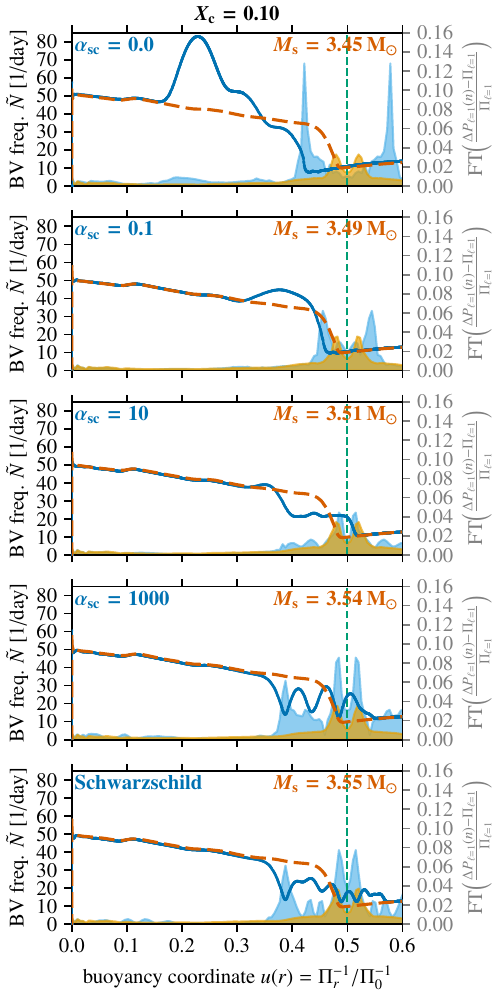}
 \caption{Same as Fig.~\ref{fig:BV_and_FT_Xc30_noos} but for $X_{\mathrm{c}} = 0.10$. The vertical dashed green line indicates the Nyquist frequency at $u(r) = 0.5$.}
 \label{fig:BV_and_FT_Xc10_noos}
\end{figure}

\begin{figure}
 \includegraphics[width=0.95\columnwidth]{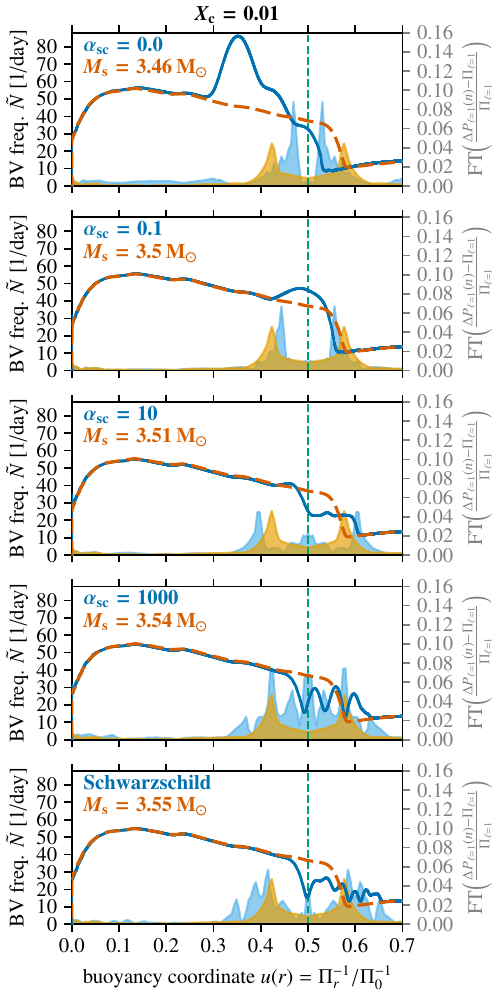}
 \caption{Same as Fig.~\ref{fig:BV_and_FT_Xc30_noos} but for $X_{\mathrm{c}} = 0.01$. The vertical dashed green line indicates the Nyquist frequency at $u(r) = 0.5$.}
 \label{fig:BV_and_FT_Xc1_noos}
\end{figure}

\bsp	
\label{lastpage}
\end{document}